\pdfoutput=1
\documentclass[fleqn,usenatbib,usedcolumn]{mnras}

\usepackage{mathptmx}
\usepackage{amsmath}
       
\usepackage[T1]{fontenc}
\usepackage{ae,aecompl}

\usepackage[british]{babel}             
     
\usepackage{graphicx}
\usepackage{url}
\usepackage{booktabs}  
\usepackage{epstopdf}  
\usepackage{microtype}  
\usepackage{float}  

\volume{{\rm in press}}

\AtBeginDocument{%
 \abovedisplayskip=3pt plus 3pt minus 9pt
 \abovedisplayshortskip=0pt plus 3pt
 \belowdisplayskip=12pt plus 3pt minus 9pt
 \belowdisplayshortskip=7pt plus 3pt minus 4pt
}

\newcommand{\Msun}{\ensuremath{\,\text{M}_\odot}}
\newcommand{\degrees}{\ensuremath{^{\circ}}}
\newcommand{\chis}{\ensuremath{\chi^{2}}}
\newcommand{\dchis}{\ensuremath{\Delta\chi^{2}}}
\newcommand{\rg}{\ensuremath{r_g}}
\newcommand{\Tdisk}{\ensuremath{T_\text{disc}}}
\newcommand{\Ndisk}{\ensuremath{N_\text{disc}}}
\newcommand{\Tbb}{\ensuremath{T_\text{bb}}}
\newcommand{\Nbb}{\ensuremath{N_\text{bb}}}
\newcommand{\TdiskP}{\ensuremath{T_\text{diskP}}}
\newcommand{\NdiskP}{\ensuremath{N_\text{diskP}}}
\newcommand{\Fscatt}{\ensuremath{f_\text{scatt}}}
\newcommand{\Eline}{\ensuremath{E_\text{line}}}
\newcommand{\Nline}{\ensuremath{N_\text{line}}}

\title[Phase-Resolved Spectroscopy of Type B QPOs]{Phase-Resolved Spectroscopy of Type B Quasi-Periodic Oscillations in GX 339--4}

\author[A.L.~Stevens and P.~Uttley]
{Abigail L.~Stevens$^1$\thanks{Email: \href{mailto:A.L.Stevens@uva.nl}{A.L.Stevens@uva.nl}} and 
Phil Uttley$^1$\thanks{Email: \href{mailto:P.Uttley@uva.nl}{P.Uttley@uva.nl}} \\
$^{1}$Anton Pannekoek Institute for Astronomy, University of Amsterdam, Science Park 904, 1098 XH Amsterdam, The Netherlands.}

\date{}

\pubyear{2016}
 
\begin{document}
\label{firstpage}
\pagerange{\pageref{firstpage}--\pageref{lastpage}}
\maketitle

\begin{abstract}
We present a new spectral-timing technique for phase-resolved spectroscopy and apply it to the low-frequency Type B quasi-periodic oscillation (QPO) from the black hole X-ray binary GX 339--4.  
We show that on the QPO time-scale the spectrum changes not only in normalisation, but also in spectral shape. 
Using several different spectral models which parameterise the blackbody and power-law components seen in the time-averaged spectrum, we find that both components are required to vary, although the fractional rms amplitude of blackbody emission is small, $\sim$\,1.4~per~cent compared to $\sim$\,25~per~cent for the power-law emission.  
However the blackbody variation leads the power-law variation by $\sim$\,0.3 in relative phase ($\sim$\,110~degrees), giving a significant break in the Fourier lag-energy spectrum that our phase-resolved spectral models are able to reproduce.  
Our results support a geometric interpretation for the QPO variations where the blackbody variation and its phase relation to the power-law are explained by quasi-periodic heating of the approaching and receding sides of the disc by a precessing Comptonising region.  
The small amplitude of blackbody variations suggests that the Comptonising region producing the QPO has a relatively large scale-height, and may be linked to the base of the jet, as has previously been suggested to explain the binary orbit inclination-dependence of Type B QPO amplitudes.
\end{abstract}

\begin{keywords}
accretion, accretion discs --
methods: data analysis --
methods: statistical --
X-rays: binaries -- 
X-rays: individual: GX 339--4
\end{keywords}


\section{Introduction}

The origin of quasi-periodic oscillations (QPOs) in X-ray binaries is still an enigma. 
They appear as Gaussian- or Lorentzian-shaped features in the averaged power spectrum of a light curve.
QPOs in black hole X-ray binaries have been observed at $0.01 - 450$\,Hz, and are broadly classified into two types, low-frequency (LF; $\sim$\,0.1 to tens of Hz) and high-frequency (HF; $\sim$\,100 Hz or more) \citep{vdKlis05, RemillardMcClintock06}. 

Due to the different timescales of variability, it is difficult to define a single mechanism responsible for the whole observed frequency range of QPOs. 
HFQPO frequencies correspond to the dynamical timescale in the inner accretion disc of the X-ray binary, which suggests that they are related to the Keplerian velocity of the accretion flow \citep{Strohmayer01, KluzniakAbramowicz01}. 
LFQPOs correspond to longer timescales, and have garnered explanations based on either a varying intrinsic luminosity or a varying emission geometry. Examples of LFQPO physical models include 
seismic oscillations in the accretion disc \citep{NowakWagoner91},
shocks in the accretion flow \citep{Chakrabarti96}, 
a precessing inner hot accretion flow \citep{StellaVietri99, IngramDoneFragile09},
nodal modulation of dynamical disc fluctuations \citep{PsaltisNorman00},
gravity waves in the accretion disc \citep{Titarchuk03}, and intrinsic variability in the base of a jet \citep{Gianniosetal04}.

In the past 10 years there has been increasing evidence suggesting that LFQPOs have a geometric origin. 
\citet{Schnittmanetal06} found that for a sample of 10 sources there is a clear correlation between binary orbit inclination and QPO amplitude, as predicted by their precessing ring model (which was motivated by the first sub-QPO-cycle spectroscopy by \citealt{MillerHoman05}). 
This result was later expanded upon for a much larger sample of observations by \citet{Mottaetal15}, who determined that the observed LFQPO amplitude has a statistically significant dependence on the orbital inclination (see also \citealt{Heiletal15b}).

Studying power spectra alone has not provided the ability to distinguish between theoretical models. 
By combining energy spectral and timing information simultaneously so that spectroscopy can probe the QPO variability timescale, we are able to look at the causal relationships between different spectral components and consider whether the QPO is caused by accretion rate fluctuations or geometric changes. 
For example, the precessing inner accretion flow model would give an observable phase relationship between the blackbody and power-law emission due to varying illumination and heating of the accretion disc by the hot inner flow.

Previous spectral-timing methods used to study QPOs such as rms spectra (e.g., \citealt{SobolewskaZycki06}, \citealt{Gaoetal14}, \citealt{AxelssonDone16}) found that the power-law component was likely responsible for the QPO emission. 
These methods give the amplitude of variability as a function of energy, but do not incorporate phase information about different energy spectral components. 
For broadband noise components, conventional cross-spectral lag studies are used to determine the phase relationship as a function of energy and frequency \citep{Uttleyetal11}. 
However, this type of signal is not appropriate for time-domain phase-resolved study, since broadband noise is not coherent on any timescale. 
For periodic and quasi-periodic variability, it is useful to attempt phase-resolved spectroscopy.

In this paper we introduce a novel spectral-timing technique for phase-resolved spectroscopy of short-timescale variability, specifically applied here to Type B LFQPOs from the black hole binary GX 339--4. 
Type B QPOs are characterised by a low level of broadband noise (correlated noise in the lowest frequencies of the power spectrum), so most of their variability power is contained in the QPO. 
This makes Type B QPOs the cleanest QPO signal for our analysis.
Our technique uses a combination of the cross-correlation function, energy spectroscopy, and simulated lag-energy spectra to find a model that can explain the data in both the energy and Fourier domains. 
In the discussion we interpret our QPO-phase-resolved spectroscopy results through the lens of a geometric precession LFQPO model, but we emphasise that this model-independent technique can be used to test any physical model that predicts spectral changes on the variability timescale. 
We conclude with a discussion of the results and their interpretation, with a forward look to how combining our new technique with data from new missions will be able to further constrain models for LFQPOs, which can constrain how matter behaves in strong gravitational fields.

\section{Data and Basic Spectral-Timing Behaviour}

GX 339--4 is a black hole candidate X-ray binary \citep{Hynesetal03} with a black hole of lower mass limit $\sim$\,7\Msun\ \citep{MunozDariasetal08} and which may possess near-maximal spin \citep{Ludlametal15}. 
The system also likely has a low binary orbit inclination ($\sim$\,40\degrees; \citealt{MunozDariasetal13}).  
The source exhibits X-ray variability on a variety of timescales. 
In the 2010 outburst \citep{Yamaokaetal10}, GX 339--4 was observed for some time in the soft intermediate spectral state, which showed strong Type B QPOs \citep{Mottaetal11}. 

\subsection{Data}
For this analysis we used data from NASA's High Energy Astrophysics Science Archive Research Center (HEASARC) taken with the Proportional Counter Array (PCA) onboard the Rossi X-ray Timing Explorer (RXTE) in 64-channel event mode with $122\,\mu$s time resolution (\texttt{E\_125us\_64M\_0\_1s}).

The following filtering criteria were used to obtain Good Time Intervals (GTIs) for analysis: Proportional Counter Unit (PCU) 2 is on, two or more PCUs are on, elevation angle $>$\,10\degrees, and target offset $<$\,0.02\degrees. 
Time since the South Atlantic Anomaly passage was not filtered on. 
The nine observation IDs with events fitting these criteria are: 95335-01-01-05, 95335-01-01-07, 95335-01-01-00, 95335-01-01-01, 95409-01-18-00, 95409-01-17-05, 95409-01-18-05, 95409-01-17-06, and 95409-01-15-06.

The data were initially binned to 7.8125\,ms time resolution ($64\,\times$ the intrinsic resolution of the data) and divided into 64\,s segments, giving 8192 time bins per segment. 
After rejecting the segments with a negative integrated noise-subtracted power (6 segments in total), there were 198 good segments of data over the nine observations, or an exposure time of 12.672\,ks. 
Note that the Type B QPO appears to sharply switch on and off on timescales of less than a few minutes (e.g., \citealt{Bellonietal05} figures A.3 and A.4), so is not expected to always contribute to the light curve, hence the segments with no detectable signal (and negative integrated power) are discarded.

\subsection{Power spectra}
\label{sec:psd}

\begin{figure}
\centering
\includegraphics[height=0.99\columnwidth,angle=90]{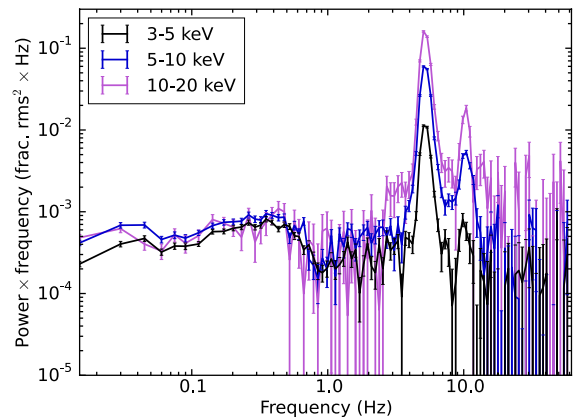}
\caption{
Power spectra in different energy bands, averaged over all available PCUs in the filtered event-mode data. 
The energy bands used are $3-5\,$keV (rms $= 7.7\%$; mean rate $= 1627.4\, \text{cts}\,\text{s}^{-1}$), $5-10\,$keV (rms $= 13.9\%$; mean rate $= 872.3\, \text{cts}\,\text{s}^{-1}$), and $10-20\,$keV (rms $= 23.1\%$; mean rate $= 183.9\, \text{cts}\,\text{s}^{-1}$).
The power spectra were geometrically re-binned in frequency with a binning factor of 1.06.
There is a Type B QPO with a centroid frequency $\nu_\text{centroid} \simeq 5.2\,$Hz and a harmonic just above $10\,$Hz. 
Individual observations have been shifted in frequency to correct for QPO centroid variations, as explained in the text. 
}
\label{fig:psd}
\end{figure}

We computed the average power spectrum in three energy bands (Figure \ref{fig:psd}) using our own code.\footnote{See Appendix A for url.} 
The power spectra were geometrically re-binned in frequency with a binning factor of 1.06. 
A power-law model of the noise level was fitted to the higher frequency power of each spectrum ($>$\,25\,Hz, by a chi-squared minimisation with a Levenberg-Marquardt algorithm) then subtracted.
There is a strong Type B QPO with a centroid frequency $\nu_\text{centroid} = 5.20\,$Hz (as determined by a chi-squared minimisation of a Gaussian profile to the average power spectrum over all energy channels) and a quality factor $Q = 6.6$ ($Q=\nu_{\text{centroid}}/\text{FWHM}$), and weak broadband noise below $1\,$Hz. 
In the literature it is common to fit QPOs with a Lorentzian profile, but since these Type B QPOs have a relatively smooth peak, they are better fit with a Gaussian.

The QPO centroid frequency shifts slightly between observations ($\nu_\text{centroid}$ ranges from 4.87\,Hz to 5.65\,Hz). 
The frequency shift smears out the QPO signal averaged over all observations, and thus reduces our signal to noise. 
To combat this, we artificially shifted the QPO centroid frequencies obtained from each RXTE ObsID to line up at the centroid of the unaltered average power spectrum ($5.20\,$Hz). 
To line up the centroid frequencies, we kept 8192 bins for each segment of data but adjusted the segment lengths, so that the width of the time bins $dt$ changed. 
The segments were adjusted by the same amount per observation. 
After adjusting, the average $dt$ over all observations is $8.153\,$ms, which gives a Nyquist frequency of $61.33\,$Hz and a Fourier-space frequency resolution of $0.015\,$Hz, and changes the exposure time to 13.224\,ks. 
This adjustment has the same effect as the shift-and-add technique commonly used when averaging multiple power spectra of neutron star kHz QPOs \citep{Mendezetal98}. 
The adjustment also corrects our phase-resolved analysis (in Section \ref{sec:technique}) to ensure that we compare relative phases, allowing for more accurate phase-resolved spectroscopy by forcing the same number of QPO cycles in one segment of data.

\subsection{Lag-energy spectrum}
\begin{figure}
\centering
\includegraphics[height=0.99\columnwidth,angle=90]{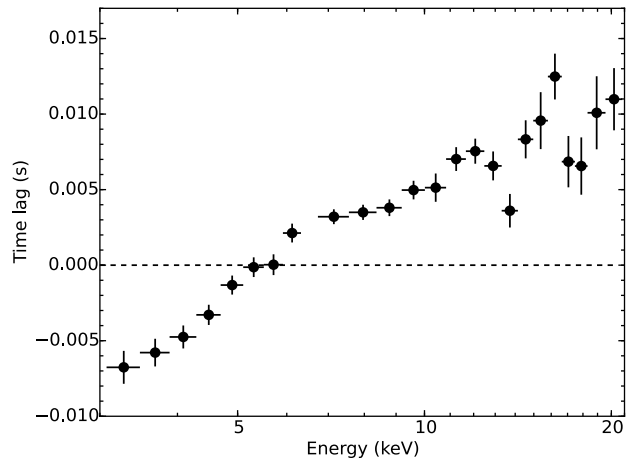}
\caption{Time lag obtained from the cross-spectrum averaged over $4-7\,$Hz, plotted versus energy. 
The dashed line indicates no time lag with respect to the variability in the reference band. 
Positive lags indicate energies which lag the reference band while negative lags lead the reference band.  
There is a clear break in the slope of the lag-energy spectrum at around 6\,keV. 
The PCA has zero counts in detector channel 11 when in event-mode, hence the gap in the data at $\sim$\,6.5\,keV.}
\label{fig:lag-energy}
\end{figure}

We also compute a lag-energy spectrum for the QPO (by averaging the cross-spectrum in the $4-7\,$Hz range), using our own lag spectrum code
\footnote{See Appendix A for url.}
following the outline in \citet{Uttleyetal14}. Using the same approach described in Section~\ref{sec:ccf}, we measure the lags for event mode energy channels from PCU2, relative to a reference band which includes the counts in the 3--20~keV range from all other available PCUs (excluding PCU2).  
The lag-energy spectrum is plotted in Figure \ref{fig:lag-energy}. 
It crosses the zero lag mark (dashed line) at the average energy of the reference band variations. 
The negative time lags denote the energy-dependent variations that lead the reference band variations, and positive time lags denote the variations that lag the reference band.
 
A flat or smooth lag-energy spectrum would indicate a simple evolution of the continuum or one spectral component over the averaged timescale. However, in Figure \ref{fig:lag-energy} there is a bump or break at $\sim$\,6\,keV. 
This indicates a more complex spectral behaviour, such as a causal relationship between separate spectral components \citep{Uttleyetal11}. 
Investigating the cause of this lag behaviour requires a more comprehensive analysis of individual energy spectra at different phases of the relative QPO cycle, or phase-resolved spectroscopy.

\section{Phase-Resolved Spectroscopy Technique} \label{sec:technique}

One of the ways to probe the physics and geometry of black hole binaries is via X-ray spectroscopy. 
Traditionally this is done by analysing the mean spectrum of a source over a set of observations, or per state of a transient outburst. 
However, even looking at spectra per observation only shows the overall spectral trends, as these timescales are much larger than those of the variability processes. 
A short segment of data would be too noisy, due to insufficient counts, for drawing detailed conclusions (see however \citealt{SkipperMcHardy16} for a more general approach using short-term spectral variability). 
The most desirable solution is to carry out sub-variability-timescale spectroscopy, or effectively phase-resolved spectroscopy for quasi-periodic signals.

For periodic signals, one can do phase-resolved spectroscopy by phase-folding the variations to stack the light curves (e.g., \citealt{Wilkinsonetal11}). 
However, this approach requires exactly periodic signals from sources with known ephemerides, which is not appropriate for QPOs (\textit{quasi}-periodic phenomena by definition). 

Previous approaches for spectral-timing of QPOs have been carried out in the time domain. \cite{MillerHoman05} used a bright source to extract spectra at maxima and minima of a QPO waveform in the broadband light curve. 
Recently, \cite{IngramvdKlis15} developed a method to reconstruct a QPO waveform for relatively-high-count-rate light curves per narrow energy band, then selecting different times in a QPO cycle for all energy bands to create and compare energy spectra. 
These methods yielded new discoveries, but are only applicable to bright sources with a well-defined QPO waveform. 
Here, we develop a more general approach that gives some of the benefits of phase-resolved spectroscopy without the requirements of a periodic signal or high count rate, using the cross-correlation function.

\subsection{Approximating phase-resolved spectra}

Consider correlated light curves defined at every time $k$ with a sinusoidal modulation at the same angular frequency $\omega$.
A narrow energy band light curve $x$ is determined for every discrete energy $E_i$ with an energy-dependent amplitude $a(E_i)$ and phase $\psi(E_i)$, while a reference band light curve $y$ for a much broader energy range has amplitude $a_{\rm ref}$ and phase $\psi_{\rm ref}$. 
Both have noise components $n(E_{i})$ and $n_{\rm ref}$, respectively.
The light curves are then expressed as:
\begin{align}
&x_k(E_i) = a(E_i) \,\sin\left(\omega\, t_k + \psi(E_i) \right) + n_{k}(E_i)\\
&y_{{\rm ref},k}= a_{\rm ref} \,\sin\left(\omega\, t_k + \psi_{\rm ref} \right) + n_{{\rm ref},k}
\end{align}
At the Fourier frequency corresponding to $\omega$, the cross spectrum $C$ with the reference band for each energy channel is:
\begin{equation}
C_{E_i,{\rm ref}} = A(E_i)\, A_{\rm ref} \,\exp{\left(i\left[\psi(E_i) - \psi_{\rm ref}\right]\right)} + C_{\rm noise}
\end{equation}
If the absolute rms-squared normalisation is used (e.g. see \citealt{Uttleyetal14}) the Fourier amplitudes $A(E_i)$ and $A_{\rm ref}$ scale with the rms values of the sine wave in the energy channel and reference band. 
Note that a similar term is produced at the corresponding negative frequency in the Fourier transform, which must also be kept for the next step.  
The noise term is produced by the products of the Fourier transforms of noise components with each other and with those of the sinusoidal signal.  

The cross-spectrum is a Fourier pair with the cross-correlation function, or CCF, so that if we inverse Fourier transform the cross-spectrum we obtain the CCF per time-delay bin $\tau_l$ for each energy channel $E_i$:
\begin{align}
\label{eqn:ccf}
CCF_{E_i,{\rm ref}}(\tau_{l}) =\,& \mathcal{F}^{-1}\left(C_{E_i,{\rm ref}}\right) \\
=\, & a(E_i)\, a_{\rm ref}\, \sin\left(\omega\,\tau_l + \left[ \psi(E_i) - \psi_{\rm ref}\right]\right)\\
& + CCF_{\rm noise}(\tau_{l})
\end{align}
After calculating the CCF, we normalise by the noise-subtracted integrated rms of the reference band, to divide out the reference band normalisation $a_{\rm ref}$. 
Thus, the CCF of the signal component is of the same form as the energy-dependent and sinusoidal original light curve of interest $x$, but with the phase defined as being relative to the reference band.  
In this way, we can recover the {\it phase-resolved} energy dependent signal.  

The CCF of the noise component is statistically independent between time segments used to calculate the CCF, and has a mean of zero.  
Thus, averaging many CCFs obtained from identical-length light curve segments can be used to suppress the noise. 
However, it is important to bear in mind that the noise in the CCF is not in general statistically independent between CCF time-delay bins, which complicates the interpretation of the errors using our approach, as discussed in the next section.

The above demonstration is shown for sinusoidal signals, whereas the Fourier pair of a QPO signal more closely resembles a damped sinusoid, so some caution needs to be applied in interpreting the energy-resolved CCF in terms of phase-resolved spectra. 
We therefore use this approach mainly as a guide to determine the best way to model the energy-dependent QPO signal in terms of variable spectral components. 
However, in Section \ref{sec:lags} we show that we can independently re-create the conventional lag-energy spectrum of the data by sinusoidally varying energy spectral parameters as inferred from our CCF method, which suggests that our approach is valid for deriving QPO phase-resolved spectra.

\subsection{Cross correlation} \label{sec:ccf}

Using the CCF we can isolate the characteristic QPO features in the time domain by correlating with a higher count-rate light curve from a broad reference energy band. 
By comparing different energy channels (``channels of interest") with the same reference band, we can connect the QPO's relative flux changes using energy channel resolution.
The CCF gives the average QPO signal of a channel of interest correlated with and relative to the reference band. 

Central to this method is establishing narrow energy channels of interest and a broad reference energy band, as outlined for cross-spectral analysis in \cite{Uttleyetal14}. 
For our RXTE PCA data, the channels of interest are taken from PCU2, as it has a well-established calibration and is switched on for the greatest number of observations.  
Since the data are filtered to have at least two PCUs on, the broad reference band comes from any other PCUs that are on (i.e. excluding PCU2, so that the reference band and channel of interest light curves are statistically independent).  
The reference band corresponds to the energy range $3-20\,$keV (absolute channels $6-48$, inclusive); this range ensures plenty of photons for optimal signal-to-noise, and fully covers the energy range used in spectral fitting in Section \ref{sec:ccf-phasespec}.

The time binning of the data is first adjusted per observation as explained in Section \ref{sec:psd}, and the cross spectrum is computed per energy channel per segment of data. 
Applying an inverse Fourier transform to each cross spectrum yields the CCF in each channel of interest per segment of data. 
The CCFs are then averaged together in the time domain per energy channel over all segments.
We then normalise the averaged CCF by $2 / (K\, \sigma_{\rm ref})$, where $\sigma_{\rm ref}$ is the absolute-normalised integrated rms of the averaged power spectrum of the reference band, and $K$ is the number of time bins per segment (as in \citealt{Uttleyetal14}). 
This gives the CCF units of count rate as deviations from the mean count rate per channel of interest, and corrects for the mixed PCUs in the reference band. 
The error is calculated for each time-delay and energy bin from the standard error on the mean CCF in that bin.

\begin{figure}
\centering
\includegraphics[height=0.99\columnwidth,angle=90]{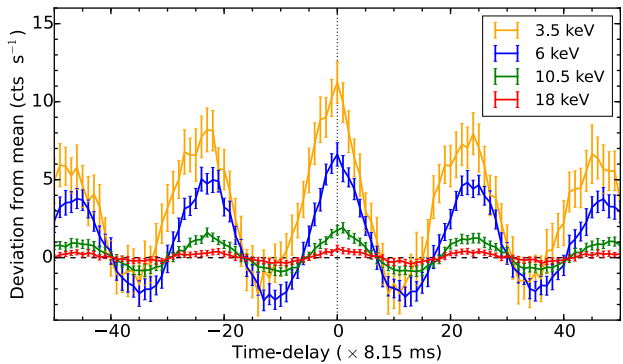}
\caption{
The CCF in four energy channels (corresponding to 3.5, 6, 10.5, and 18 keV), normalised to have units of count rate. The CCF has a larger amplitude in the lower energy channels, but the lower energy channels also have higher mean count rates.
}
\label{fig:multiccf}
\end{figure}

The CCF in four energy channels is shown in Figure \ref{fig:multiccf}. 
Since cross-correlation phase-locks the features of the signal in the channels of interest to the signal in the reference band, it produces a relatively strong signal modulation, even with a relatively low count rate in each energy channel. 

This CCF phase-resolved spectroscopy technique does not make assumptions about an underlying QPO waveform, and is applicable independent of the QPO emission mechanism. 
Thus it is a powerful method to test physical QPO models that predict specific spectral-timing behaviour. 
Since the reference band and channels of interest overlap in energy space and are taken simultaneously from independent detectors, we assume that if a signal is present in the light curves, there will be some degree of correlation between the channels of interest and the reference band. 
The Poisson noise components are uncorrelated, so the resulting CCF contains the (correlated) QPO signal, with less noise than the input light curves. 
Our technique yields a count rate per energy channel in each time-delay bin, as deviations from the mean count rate per energy channel. 
After adding the CCF to the mean count rate per energy channel, these effectively-phase-resolved spectra can be loaded into a spectral fitting program like \textsc{xspec}.

As already briefly mentioned in the previous section, statistical fluctuations between CCF time-delay bins for a given energy channel are correlated with one another.  
This effect arises from the component of the noise resulting from the multiplication of the Poisson noise by the intrinsic signal, resulting in `structure' in the noise across phase bins, the strength of which depends on the shape of the intrinsic signal power spectrum (see \citealt{Bartlett55,BoxJenkins76} for a detailed description of the errors in the CCF associated with autocorrelated processes). 
Thus the phase-resolved spectra produced by our method are not strictly independent in time, although they are independent in energy (since different energy bins have independent Poisson noise terms).  
Therefore one cannot easily interpret the results of conventional fitting techniques, such as \chis, which are used to simultaneously fit the spectra from multiple phase bins. 
To compensate, we will implement bootstrapping (Section \ref{sec:bootstrapping}) and lag-energy spectral comparison (Section \ref{sec:lags}) to determine accurate errors on our description of the modulation of spectral parameter values and check the appropriateness of the spectral model.

\section{Results}

\subsection{Energy-dependent CCF}

\begin{figure}
\centering
\includegraphics[height=0.99\columnwidth,angle=90]{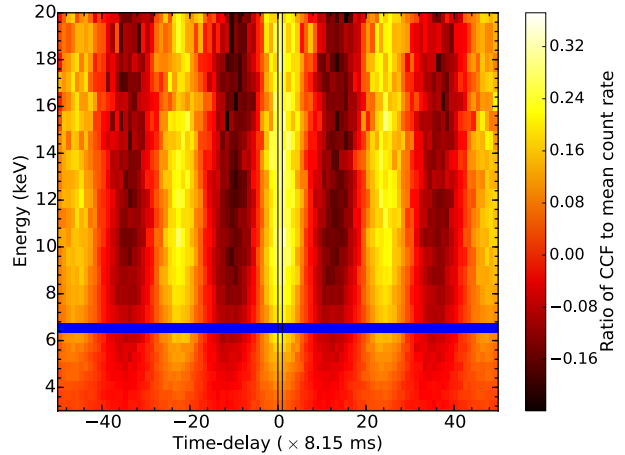}
\caption{Cross-correlation function (CCF) plotted in two dimensions. 
The colour map shows the ratio of the CCF amplitude in each bin to the mean count rate of the energy channel. 
The zero time-delay bin is indicated by the solid black outline. 
The PCA has zero counts in detector channel 11 when in event-mode, hence the gap in the data at $\sim$6.5\,keV as indicated by the blue bar.
}
\label{fig:2dccf}
\end{figure}

The CCF can be plotted in a two-dimensional colour plot (Figure \ref{fig:2dccf}), to show the simultaneous time variability and energy spectra. 
The colour map shows the ratio of the CCF amplitude to the mean count rate of each energy channel, because the lower energy channels have larger CCF amplitudes but also higher mean count rates. 
The energy range shown along the $y$-axis, $3-20\,$keV, is the same range used for the reference band in Section \ref{sec:technique} and for spectral fitting in Section \ref{sec:ccf-phasespec}. 
Although the mean count rates are higher at lower energies, the amplitude of the QPO signal in the CCF is largest at $\sim$10\,keV. 
The CCF of the QPO has the approximate shape of a damped sinusoid in each energy channel, so henceforth we will refer to a QPO phase in the same way that one can consider a sinusoidal phase. 
As noted in Section \ref{sec:psd}, we correct for variations in the QPO centroid frequency between observations by artificially adjusting the length of the light curve segments, such that the same number of QPO cycles are obtained in each data segment used to calculate the CCF.  
In doing this and then interpreting the results as a phase-resolved spectrum, we implicitly assume that time-delays between energy bands are constant in phase and not constant in time, so that the intrinsic phase relationships between energy bands are preserved and not blurred out by the adjustment process.  
This approach will be borne out by the success of our phase-resolved spectral model in reproducing the shape of the lag-energy spectrum, whether it is adjusted for QPO frequency changes or not (see Section~\ref{sec:lags}).

\subsection{Time-averaged energy spectrum}

\begin{figure}
\centering
\includegraphics[height=0.99\columnwidth, angle=270]{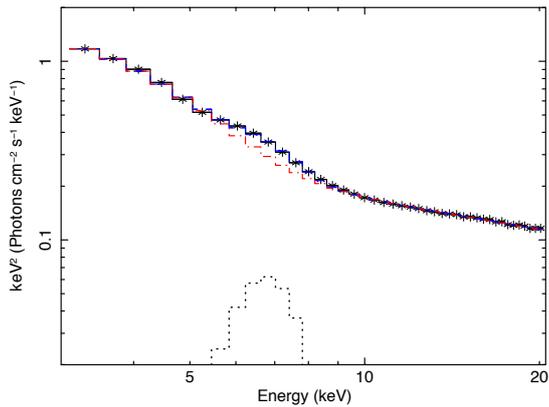}
\caption{The time-averaged spectrum of the data in Standard-2 data mode, unfolded through the instrument response matrix and fit with the model \textsc{phabs}\,$\times$\,\textsc{(simpl\,*\,diskbb\,+\,gauss)}. The solid black line with star points represents the data, the dashed blue line represents the total model, the dash-dotted red line is \textsc{simpl} convolved with \textsc{diskbb} (since it is a convolution model, there is only one output curve), and the dotted line is \textsc{gauss} representing the iron K$\alpha$ emission line. This is a simple phenomenological spectral fit, with $\chis/\,$d.o.f.$\,= 229.2/37$.
}
\label{fig:std2-spec}
\end{figure}

The time-averaged spectrum obtained from the RXTE PCA Standard-2 data mode, shown in Figure \ref{fig:std2-spec}, was used for identifying the general energy spectral features and considering which spectral models to fit for the phase-resolved data. 
This data mode has finer energy binning than the high time resolution event-mode data used for the rest of the analysis.

The spectrum shows a soft blackbody-like component, a hard power-law component, an iron K$\alpha$ emission line, and neutral hydrogen absorption. We fit this with \textsc{phabs}\,$\times$\,\textsc{(simpl\,*\,diskbb\,+\,gauss)}, a simple phenomenological model containing the key components. 
While the presence of the iron line indicates that a broad reflection continuum should also be included, we expect that there would be degeneracies between an upscattered power-law and the reflection components, particularly for the coarser energy binning of event-mode spectra. 
Since our main interest was the possible variability of the soft and hard spectral components and any phase delays between them, any reflection variability was accounted for with the power-law and Gaussian components. 
As later noted, we find that the Gaussian iron line is not required by our fits to vary on the QPO timescale so we do not expect variable reflection to contaminate the inferred variation of other spectral components.

\subsection{CCF phase-resolved spectroscopy}
\label{sec:ccf-phasespec}

To carry out phase-resolved spectroscopy using the energy-dependent CCF, we select energy spectra at each time-delay bin of the CCF covering one QPO cycle. 
Since we focus on relative phases instead of absolute phases for simplicity, we define the QPO phase to be zero when the CCF in channel 15 is at a minimum; this is in time-delay bin -10. 
By selecting until the next minimum in channel 15 at time-delay bin +13, we have 24 time-delay bins in total for one QPO cycle. 
This time range of data covers the largest maximum and minimum in CCF amplitude so that we can see the most prominent differences in energy spectra. 
The instrument response matrix uses the calibration of PCU 2, since the channels of interest are taken from that detector.

\begin{figure} 
\centering
\includegraphics[height=0.99\columnwidth, angle=270]{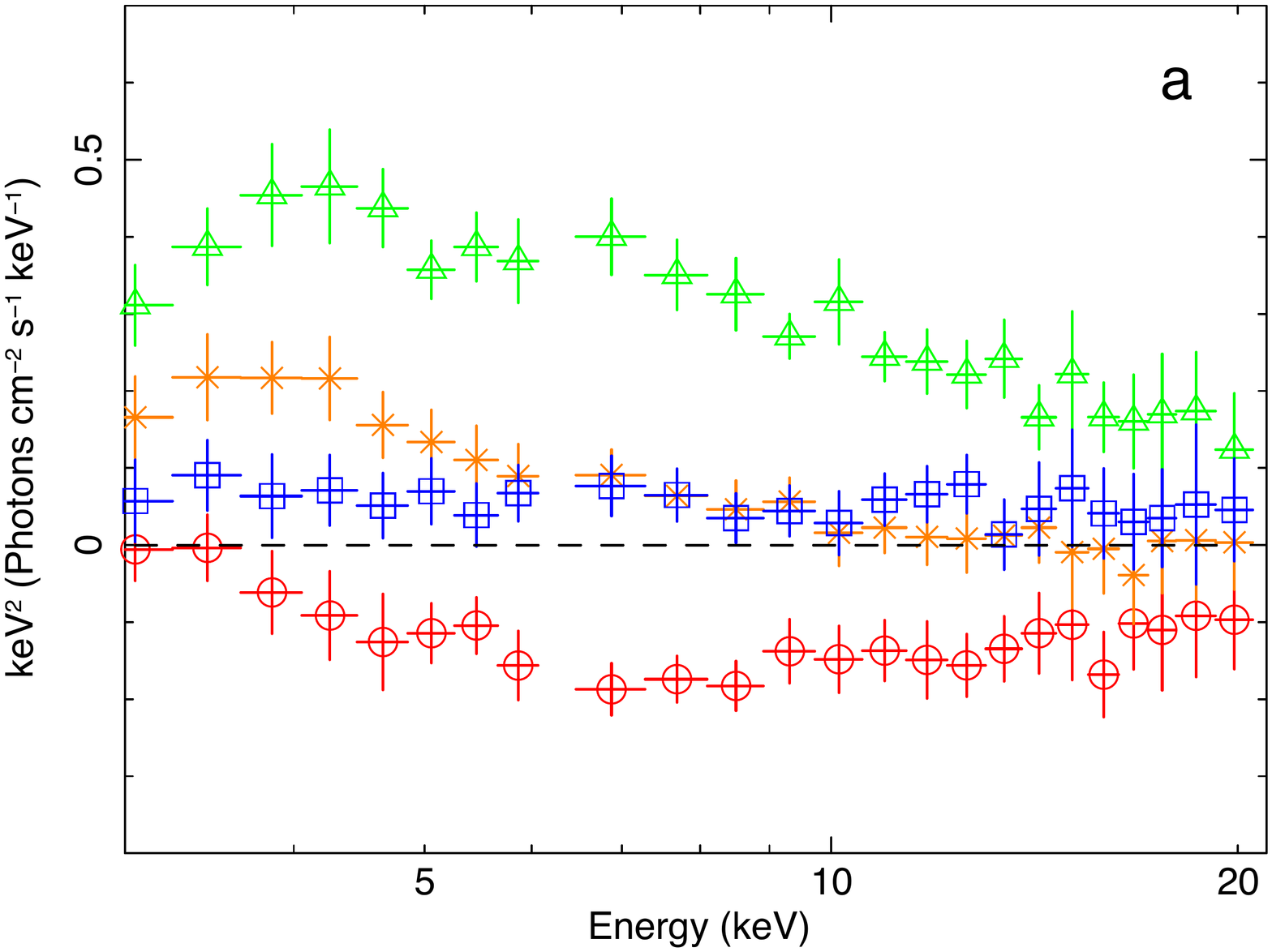}
\includegraphics[height=0.99\columnwidth, angle=270]{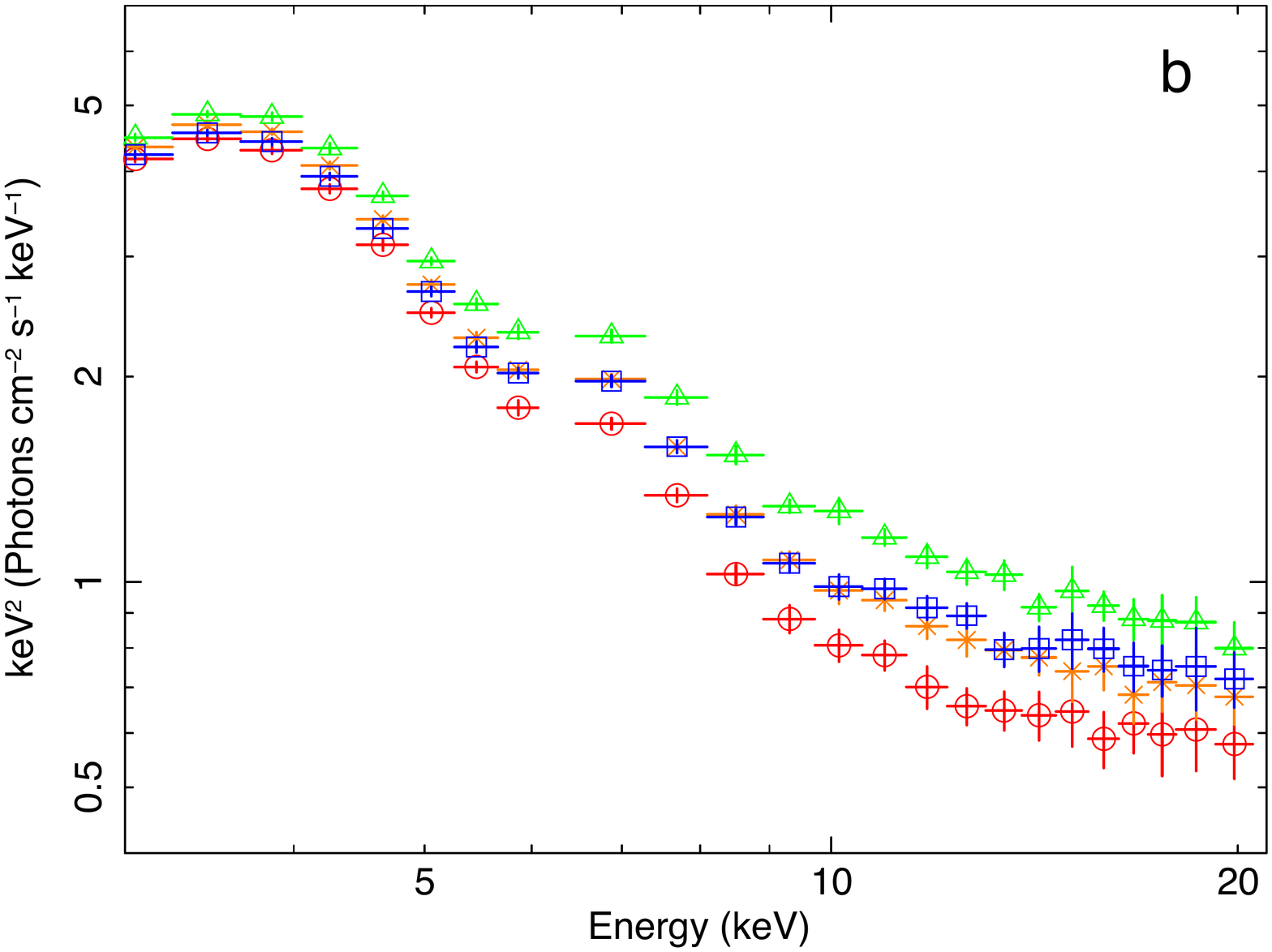}
\caption{\textbf{a:} Deviations from the mean spectrum at four QPO phases discussed in the text. 
The black dashed line denotes zero deviation from the mean spectrum.
The red circles are for a QPO phase of 0\degrees, orange X's for 90\degrees, green triangles for 180\degrees, and blue squares for 270\degrees. 
The PCA has zero counts in detector channel 11 when in event-mode, hence the gap in the data at $\sim$6.5\,keV.
It is evident that the shape of the spectrum, in addition to the normalisation, changes with QPO phase.
\textbf{b:} Energy spectra at the same four QPO phases, generated by adding the above deviations to the mean spectrum. Data points have the same meaning as in the top panel. For visualisation in both plots, the spectra have been divided by the instrument's effective area as mentioned in the text.
}
\label{fig:ccf-spec}
\end{figure}

Figure \ref{fig:ccf-spec}a shows the deviations from the mean spectrum at four QPO phases, plotted in $E\,F(E)$ units. 
Figure \ref{fig:ccf-spec}b shows the total energy spectra at the same four QPO phases, which are the deviations shown in the upper panel added to the mean spectrum.
For visualisation here, the energy spectra have been ``fluxed", i.e., divided by the effective area of the instrument response (as in \citealt{Vaughanetal11}). In this figure, the QPO phase of $\sim$\,0\degrees is at time-delay bin -10, $\sim$\,90\degrees is at time-delay bin -5, $\sim$\,180\degrees is at time-delay bin +1, and $\sim$\,270\degrees is at time-delay bin +6. 
Time-delay bins within one QPO cycle will be referred to as phase bins.  
In the remainder of this paper, we only focus on the spectral-parameterisation of variations during the central, highest-amplitude cycle seen in the CCF.  
However, we note here for completeness that an analysis of additional cycles yields consistent parameter variations (with similar phase relations to the strongest cycle), albeit with decaying amplitudes of variability, as expected due to the damped sinusoidal shape of the CCF.

For the full phase-resolved spectroscopy, we simultaneously fit energy spectra from time-delay bins -10 to +13 (inclusive, and including 0), giving 24 spectra in total; the 5.2\,Hz centroid frequency corresponds to a timescale of 0.192\,s, so the 8.153\,ms time binning gives 23.5 phase bins per QPO cycle (rounded to 24, to give complete coverage of a QPO cycle). 
Our aim is to obtain the simplest possible parameterisation of the QPO spectral variability in terms of a spectral model with variable components.  
We therefore systematically untied and froze different combinations of parameters to find an optimal fit without over-fitting, such that the resulting model makes physical sense.
We tested spectral models which all include constant neutral absorption, a soft blackbody component, a hard power-law component and an iron K$\alpha$ line. 
These components are clearly required by the time-averaged spectrum (Figure \ref{fig:std2-spec}).

The power-law is likely to be produced by Compton-upscattered disc photons. To model this component, we use \textsc{simpl}. 
\textsc{simpl} is a convolution model intended as a simple parameterisation of Comptonisation from Compton-thin plasmas by scattering a fraction (\Fscatt) of the input seed spectrum to make a power-law of photon index $\Gamma$~\citep{simpl}. 
For a given, constant seed spectrum, \Fscatt\ is thus a proxy for the normalisation of the power-law component.  
In this spectral model, the power-law lower-energy cutoff follows the low energy shape of the seed spectrum, which in our case is a multi-colour disc blackbody. 
\textsc{simpl} outputs the sum of the scattered power-law and the unscattered seed photon spectrum, so it contains a built-in anticorrelation between the normalisation of the power-law and the normalisation of the observed part of the seed spectrum. 
To remove this effect, we created a new model \textsc{simpler} where the output seed spectrum is the same as the input. 
This would correspond to the case where the Compton scatterer does not obscure the observer's view of the seed spectrum, as would be expected for an inner hot flow that is physically separate from the accretion disc. 
As is the default for \textsc{simpl}, only upscattering is accounted for (indicated by freezing the model scattering switch $U=1$ in Tables \ref{tab:1BB-bestvals}--\ref{tab:pBB-bestvals}).  

Using \textsc{xspec} version 12.8.2 \citep{XSPEC} we fitted the phase-resolved energy spectra in the energy range $3 - 20\,$keV, ignoring detector channel 11 (which has zero counts in RXTE PCA event-mode).
The solar abundance table was set to \texttt{vern} \citep{Verneretal96} and the photoionization cross-section table was set to \texttt{wilm} \citep{Wilmsetal00}. 
The systematic error was set to 0.5~per~cent.

The models we tested are \textsc{phabs}$\times$\textsc{(simpler*diskbb+gauss)} (``model 1''), \textsc{phabs}$\times$\textsc{(simpler*diskbb+bbodyrad+gauss)} (``model 2''), and \textsc{phabs}$\times$\textsc{(simpler*diskpbb+gauss)} (``model 3''); they are summarised with their best parameterisations in Table \ref{tab:specmodels}. 
These models represent the simplest possible system geometries: a power-law-emitting region and a blackbody-emitting accretion disc; 
a power-law emitting region, a blackbody-emitting accretion disc, and a secondary blackbody from the power-law heating a smaller, single-temperature region; 
and a power-law-emitting region and a blackbody-emitting accretion disc with a different (compared to the standard disc) and possibly varying radial dependence of temperature.
Details about the fits and motivations are given in subsections \ref{sec:spec-model-1}, \ref{sec:spec-model-2}, and \ref{sec:spec-model-3}, respectively. The neutral hydrogen column density was frozen to $N_H = 6\times10^{21}\,\text{cm}^{-2}$ \citep{ReynoldsMiller13}.

For carrying out the QPO phase-resolved spectroscopy, we began with every spectral parameter tied across QPO phase. 
We then systematically stepped through the parameters, untying each individually so that it can vary across phase, and assessing the fit. 
If untying a parameter gave a lower $\chis$ and physically reasonable parameter values, it remained untied and we investigated whether a second parameter should be untied. 
This continued until there was no significant improvement in $\chis$ and/or the parameter values became physically unreasonable. 

As mentioned in Section \ref{sec:ccf}, although the count rates in energy bins within a single phase-resolved spectrum are statistically independent, the phase-resolved spectra are not statistically independent from each other, in that there is some correlation in errors between different phases of the same energy bin.  
Therefore the \chis\ fit statistic returned by fitting the phase-resolved spectra together does not have the same meaning as a conventional \chis\ statistic.  
The correlation between phase bins effectively means that ratio between the \chis\ and degrees of freedom is reduced, so we use the following procedure only as a guide to find the best parameterisation of each model, rather than using an explicit goodness-of-fit test.  
We found during our fits that the plausible best-fitting models converged to $\chis/\text{d.o.f.}\sim130/500$. 
Assuming that this \chis\ corresponds to a reduced $\chis \sim1$, a significant improvement in the fit at the 3-$\sigma$ level should correspond to a $\Delta\chis$ of $46\times 130/500 \simeq 12$, where 46 is the formal $\Delta\chis$ improvement corresponding to a 3-$\sigma$ significance improvement in a nested model for 23 additional free parameters, i.e. equivalent to freeing a spectral model parameter in our simultaneous fit of 24 spectra.  
In Sections \ref{sec:spec-model-1}, \ref{sec:spec-model-2}, and \ref{sec:spec-model-3} we use this $\Delta\chis > 12$ criterion for a significant improvement to investigate which spectral parameters vary with QPO phase.  
However, in view of the fact that our approach is merely a guide to search for a spectral model parameterisation of the phase-dependent QPO variability, and is not statistically rigorous, in Sections \ref{sec:bootstrapping} and \ref{sec:lags} we assess the statistical significance of the parameter variations and their phase lags with bootstrapping and direct comparison of our best-fitting QPO spectral models with the lag-energy spectrum.

\begin{table*} 
\centering
\begin{tabular}{c l l l l c c }
\toprule
\# & Energy Spectral Model & Untied & \chis\,/\,d.o.f.\ & $ \phi_{FS} - \phi_\Gamma$ & $\phi_{FS} - \phi_{BB}$ & Lag \chis\,/\,d.o.f.\  \\
\midrule
1 & \textsc{phabs}\,$\times$\,\textsc{(simpler\,*\,diskbb\,+\,gauss)} &  \Fscatt, $\Gamma$, \Tdisk & $143.9 / 502$ & $0.01 \pm 0.02$ & $0.32 \pm 0.02$ &  $37.9/24$  \\
2 & \textsc{phabs}\,$\times$\,\textsc{(simpler\,*\,diskbb\,+\,bbodyrad\,+\,gauss)} &  \Fscatt, $\Gamma$, $T_{\text{bb}}$ & $128.4 / 500$ & $0.019 \pm 0.009$ & $0.28 \pm 0.02$ & $45.5/24$ \\ 
3 & \textsc{phabs}\,$\times$\,\textsc{(simpler\,*\,diskpbb\,+\,gauss)} & \Fscatt, $\Gamma$, $p$ & $126.8 / 501$ & $0.02 \pm 0.01$ & $0.10 \pm 0.04$ & $47.2/24$  \\
\bottomrule
\end{tabular}
\caption{ A summary of the best parameterisation for each spectral model tested. 
The first column gives the number of each energy spectral model, for reference throughout the paper, and the second column states the models. 
The ``Untied'' column indicates which model parameters are untied between phase-resolved spectra to give the best parameterisation (see Sections \ref{sec:spec-model-1}-\ref{sec:spec-model-3}). 
The next column gives the \chis\ fit statistic for the energy spectral fits of each parameterisation (assuming the data are statistically independent, which they are not; see Section \ref{sec:bootstrapping}). 
The columns ``$\phi_{FS} - \phi_\Gamma$'' and ``$\phi_{FS} - \phi_{BB}$'' give the normalised phase difference in the parameter variations between an untied \Fscatt\ and untied $\Gamma$, and an untied \Fscatt\ and an untied blackbody parameter (see Section \ref{sec:dphase}).
The column ``Lag \chis\,/\,d.o.f.'' gives the fit statistics for lag-energy spectrum simulated from the energy spectral model and fitted to the data (see Section \ref{sec:lags} and Figure \ref{fig:lag-energy-comparison}). 
}
\label{tab:specmodels}
\end{table*}

\subsubsection{Spectral model 1} \label{sec:spec-model-1}

The first energy spectral model tested was \textsc{phabs}\,$\times$\,\textsc{(simpler\,*\,diskbb\,+\,gauss)}. 
This is the simplest possible model to explain the features in Figures \ref{fig:std2-spec} and \ref{fig:ccf-spec}, where the blackbody component is intended to capture all possible soft blackbody emission. 
To test which parameterisation best fits the 24 sequential energy spectra from one QPO cycle, we began with all parameters tied (and $N_\text{H}$ frozen and the upscattering-only flag set), which gave $\chis=7114.5$ for 569 d.o.f. 
For all fits, the power-law photon index $\Gamma$ was allowed values in the range $1.0-3.1$, disc maximum temperature \Tdisk\ was allowed values in the range $0.5-1.0$\,keV, the Gaussian line centroid energy \Eline\ was allowed $5.5-7.0$\,keV, and all other parameters were allowed the full range of values (unless frozen). 

We then untied each parameter one at a time to see which was required to vary given our simple \dchis\ criterion. 
Untying the scattering fraction \Fscatt\ gave a significant improvement to the fit with $\chis=393.0$ for 546 d.o.f. 
This is also physically motivated: untying \Tdisk, the disc blackbody normalisation \Ndisk, or any Gaussian line parameters did not account for the large hard continuum variations clearly seen in Figures \ref{fig:2dccf} and \ref{fig:ccf-spec}. 
By next untying $\Gamma$, we saw a large improvement in the fit to $\chis=176.5$ for 523 d.o.f. 
This changing power-law slope is also visibly evident in Figure \ref{fig:ccf-spec}a when comparing phases 0\degrees\ and 180\degrees\ (the red and green points) above $\sim6.5$\,keV.
A third parameter that significantly improved the fit when allowed to vary with phase is \Tdisk, which when untied gave $\chis=141.4$ for 500 d.o.f. 
Variations in the blackbody component are required to account for the variations of the lower-energy `soft excess' feature, seemingly out of phase with the harder power-law continuum, which are seen in Figure \ref{fig:ccf-spec}a. 
Untying \Ndisk, \Eline, or $\sigma_\text{line}$ as the third varying parameter returned worse fits.
An untied Gaussian normalisation \Nline\ yielded a smaller $\chis\,/\,$d.o.f. ($132.8/500$) than an untied \Tdisk, although the improvement was not formally accepted according to our significance criterion. 
However, examining the returned parameter values and resultant spectra showed that the broadened iron line component was subsuming the role of the higher temperature end of the blackbody emission. 
The fit pushed \Eline\ to the lowest value of the allowed range ($5.5\,$keV) and gave the line width $\sigma_\text{line} > 1\,$keV, which we believe is unphysical. 
To mitigate this degeneracy, we froze $\sigma_\text{line} = 0.97$ in the final parameterisation. 
The value of $\sigma_\text{line}$ was found using the \textsc{xspec} \texttt{steppar} command on the mean event-mode spectrum; the best-fitting value was $\sigma_\text{line} = 1.0$, so to push the value of $\sigma_\text{line}$ away from the boundary we decided on the lowest value while keeping within the 1-$\sigma$ fit region, $\sigma_\text{line} = 0.97$. 
In general, the smaller the width of the iron line, the more physically consistent the model fit is. 
Untying a fourth parameter was not able to both significantly improve the fit and return physically reasonable parameter values.

While the fits somewhat preferred untying the disc blackbody temperature to the disc blackbody normalisation, the two parameters were highly degenerate. 
To disentangle this degeneracy and ensure that any temperature variations in the bootstrapping were not due to a different normalisation, \Ndisk\ was frozen at $2505.72$ (the best-fitting value). 

\begin{table} 
\centering
\begin{tabular}{l l l l}
\toprule
Component & Parameter & Value & Notes \\
\midrule
\textsc{phabs} & $N_\text{H}$ ($\times 10^{21}$~cm$^{-2}$) & 6.0 & Frozen \\
\textsc{simpler} & $\Gamma$ & $2.52\,\dagger$ & Untied \\
\textsc{simpler} & \Fscatt & $0.156\,\dagger$ & Untied \\
\textsc{simpler} & $U$ & 1 & Frozen \\
\textsc{diskbb} & \Tdisk (keV) & $0.8307\,\dagger$ & Untied \\
\textsc{diskbb} & \Ndisk & 2505.72 & Frozen \\
\textsc{gauss} & \Eline (keV) & $6.28\pm0.03$ & Tied \\
\textsc{gauss} & $\sigma_\text{line}$ (keV) & 0.97 & Frozen \\
\textsc{gauss} & \Nline & $2.28\pm0.08\times10^{-2}$ & Tied \\
\bottomrule
\end{tabular}
\caption{Values of the best-fitting parameterisation for model 1 as explained in Section \ref{sec:spec-model-1}. Mean values are listed for the untied parameters, indicated by $\dagger$. Errors on the tied parameters were computed for the 90\% confidence interval with \textsc{xspec}'s MCMC error routine. 
\textsc{xspec} gives a fit statistic of $\chis=143.9$ for 502 d.o.f.
\label{tab:1BB-bestvals}
}
\end{table}

\begin{figure} 
\centering
\includegraphics[height=0.99\columnwidth,angle=90]{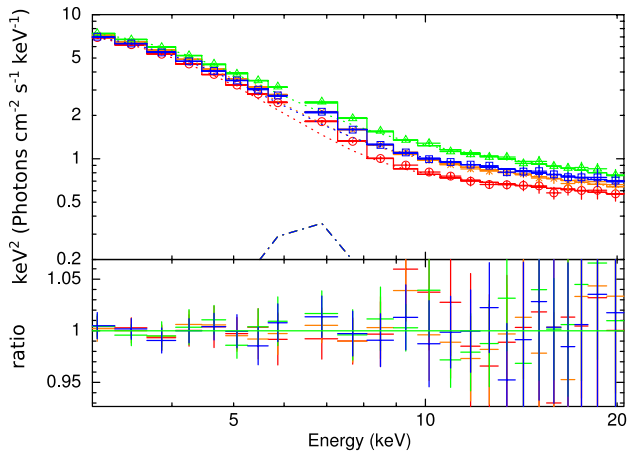}
\caption{ 
Phase-resolved energy spectra with model fits at four relative QPO phases, unfolded with the instrument response matrix and plotted in $E F(E)$ units. The colours have the same meaning as in Figure \ref{fig:ccf-spec}. 
The ratio of the data to the model is shown below the spectra.
Since the iron line values are frozen or tied, the iron line components of each spectrum stack perfectly. 
The PCA has zero counts in detector channel 11 when in event-mode, hence the gap in the data at $\sim$6.5\,keV.
}
\label{fig:1BB-4phasefit}
\end{figure}

\begin{figure}
\centering
\includegraphics[width=0.99\columnwidth]{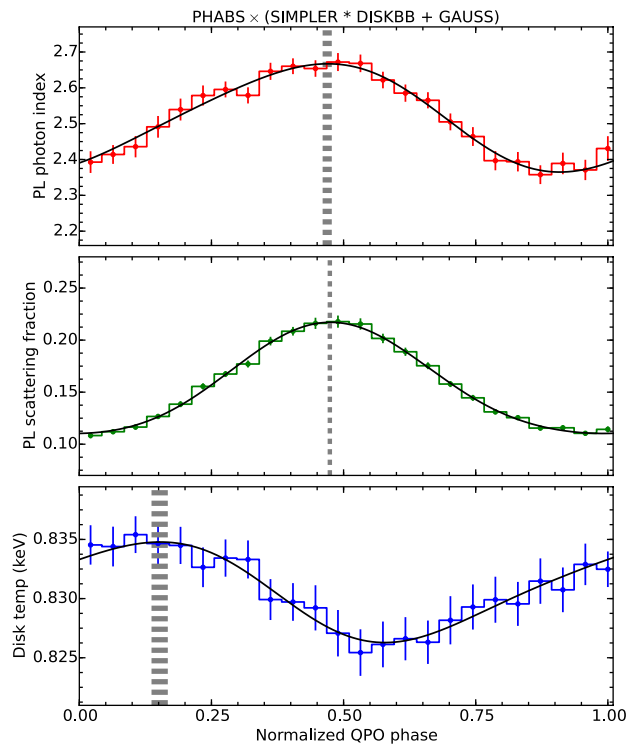}
\caption{ 
Best-fit values for untied spectral parameters at each phase in a QPO cycle for spectral model 1.  
The (1-$\sigma$) error bars were determined using bootstrapping (see Section~\ref{sec:bootstrapping}).  
The top panel shows $\Gamma$ values in red, the middle panel shows \Fscatt\ values in green, and the bottom panel shows \Tdisk\ values in blue. 
The error bars represent $1 \sigma$ errors (see Section \ref{sec:bootstrapping} for details).
The fractional rms amplitude variation of $\Gamma$ is 4.2 per cent, of \Fscatt\ is 24 per cent, and of the blackbody flux is 1.4 per cent (from luminosity $\propto T^4$ with constant blackbody normalisation).
The best-fitting function (Equation \ref{eqn:fitfunc}) of each parameter is shown in solid black (the parameters of the best-fit function for the untied parameters in each spectral model are listed in Table \ref{tab:fitfuncval}). 
The phase of the maximum of each fit function is marked with a dashed gray line, the thickness of which represents the 1-$\sigma$ error. The normalised phase difference between \Fscatt\ and $\Gamma$ is $0.01 \pm 0.02$, and between \Fscatt\ and \Tdisk\ is $0.32 \pm 0.02$; this is discussed further in Section \ref{sec:dphase}.
}
\label{fig:parvar-1BB}
\end{figure}

The best parameterisation for model 1 has \Fscatt, $\Gamma$, and \Tdisk\ untied, and \Ndisk\ and $\sigma_\text{line}$  frozen. 
The value of each parameter is listed in Table \ref{tab:1BB-bestvals}, with the mean value quoted for the untied parameters. 
Note that the \chis\,/\,d.o.f. in the table caption corresponds to the fit with \Ndisk\ and $\sigma_\text{line}$ frozen.
The same four spectra shown in Figure \ref{fig:ccf-spec} are also shown in Figure \ref{fig:1BB-4phasefit}, unfolded through the instrument response matrix and through models with the best-fitting spectral parameter values. The value of each untied parameter per phase bin is shown in Figure \ref{fig:parvar-1BB}. 
The errors on the values of the untied spectral parameters in each bin, and on the phase of their variations, were determined via bootstrapping the data (see Section \ref{sec:bootstrapping}). 
The phase differences between the untied parameter variations are discussed in Section \ref{sec:dphase}.
The equivalent width of the iron line varies with the opposite phase of the power-law continuum during a QPO cycle in the range $0.35-0.48$\,keV. Note that since the iron line flux is constant, these equivalent width variations are due to the observed power-law continuum variations (which as we later show are likely due to geometric, not intrinsic, changes). Therefore, in this interpretation, the varying equivalent width does not represent a changing intrinsic disc reflection, but we report it here for the sake of completeness.

As an extension of model 1, we considered a variant on the \textsc{diskbb} model with a free spectral hardening factor \citep{ShimuraTakahara95} that changes the apparent peak temperature of the disc without changing the integrated disc flux, to represent variable scattering in the disc atmosphere. 
Our fits showed that the data also required a variation of \textsc{diskbb} normalisation and/or temperature, i.e. the disc flux is required to change and hardening factor variations alone cannot explain the modulation of the blackbody emission.  
Effectively, hardening subsumes only part of the effect of an intrinsic temperature change (the change in spectral shape), without producing variations of disc flux which are also required by the data.  
As these results did not contribute any additional interpretation to the data beyond that of model 1, they are not reported further here.

\subsubsection{Spectral model 2} \label{sec:spec-model-2}

In the final best-fitting parameterisation of model 1, \Tdisk\ was required to vary, but the values had a very small variation ($0.8254-0.8354\,$keV). 
For the model component \textsc{diskbb}, which has a fixed radial temperature profile, this variation implies that the whole disc blackbody is changing in effective temperature. 
It seems more physically plausible that a patch of the disc is varying, e.g. due to heating by the varying power-law, while the intrinsic disc emission remains constant (particularly since the temperature variation is so small).
To model the temperature variations required by this model parameterisation, we next tested a spectral model with an added single-temperature blackbody component \textsc{bbodyrad} that is allowed to vary between phase bins, while both the \textsc{diskbb} normalisation and temperature are tied across phase. 
Following that, in the next subsection, we tested a varying radial temperature dependence with the disc blackbody model \textsc{diskpbb} (spectral model 3).

Our spectral model 2, incorporating an additional single-temperature blackbody, is defined to be \textsc{phabs}$\,\times$\,\textsc{(simpler\,*\,diskbb\,+\,bbodyrad\,+\,gauss)}. 
To determine the most effective parameterisation we followed the same procedure as in the previous section. 
For all fits, \Tdisk\ was allowed $0.6-1.0$\,keV, $T_{\text{bb}}$ was allowed $0.1-1.0$\,keV, and all other parameters were allowed the same range of values as in model 1 (unless frozen). 

\begin{figure}
\centering
\includegraphics[width=0.99\columnwidth]{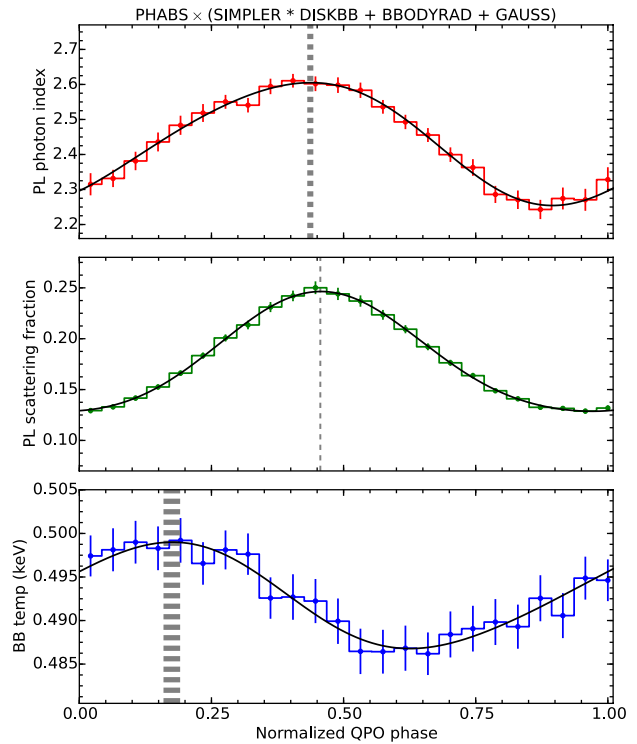}
\caption{Same meaning as Figure \ref{fig:parvar-1BB}, for model 2.  
The top panel shows $\Gamma$ parameter values in red, the middle panel shows \Fscatt\ parameter values in green, and the bottom panel shows \Tbb\ parameter values in blue. The fractional rms amplitude variation of $\Gamma$ is 4.9 per cent, of \Fscatt\ is 23.5 per cent, and of the blackbody flux is 3.5 per cent (or 1.4 per cent in total blackbody flux, allowing for the constant disc blackbody component).
The normalised phase difference between \Fscatt\ and $\Gamma$ is $0.019 \pm 0.009$, and between \Fscatt\ and \Tbb\ is $0.28 \pm 0.02$. \label{fig:parvar-2BB}}
\end{figure}

Starting with all parameters tied across QPO phase (and $N_\text{H}$ frozen), we found $\chis=7079.5$ for 567 d.o.f. 
As with model 1, the first and second spectral parameters that needed to be untied were \Fscatt\ ($\chis=392.2$ for 544 d.o.f.) and $\Gamma$ ($\chis=157.5$ for 521 d.o.f.).
By then untying \Tbb\ we see further improvement of the fit, to $\chis=125.7$ for 498 d.o.f. 
Again, an untied \Nline\ gave a slightly better statistical fit ($\chis$\,/\,d.o.f.\,$=123.3/498$) than an untied \Tbb, but it also returned an implausibly low value for the line energy. 
So, we selected the untied \Tbb\ as the best fit for this set and froze $\sigma_\text{line}=0.82$ (the best-fitting value found using \texttt{steppar} on the mean event-mode spectrum) in the final parameterisation.
The fits slightly preferred untying \Tbb\ to untying \Nbb. 
In order to break the degeneracy between \Tbb\ and \Nbb, we froze \Nbb\ at 8857.68 (the best-fitting value).
The best parameterisation for model 2 and the values of each parameter are stated in Table \ref{tab:2BB-bestvals}. 
The value of the untied parameters at each phase bin is shown in Figure \ref{fig:parvar-2BB}. 
The iron line equivalent width varied in the range $0.23 - 0.32$\,keV with the opposite phase of the power-law variations, due to the varying power-law continuum.

\begin{table} 
\centering
\begin{tabular}{l l l l}
\toprule
Component & Parameter & Value & Notes \\
\midrule
\textsc{phabs} & $N_\text{H}$ ($\times 10^{21}$~cm$^{-2}$) & 6.0 & Frozen \\
\textsc{simpler} & $\Gamma$ & $2.44\,\dagger$ & Untied \\
\textsc{simpler} & \Fscatt & $0.179\,\dagger$ & Untied \\
\textsc{simpler} & $U$ & 1 & Frozen \\
\textsc{diskbb} & \Tdisk (keV) & $0.94\pm0.01$ & Tied \\
\textsc{diskbb} & \Ndisk & $1090\pm90$ & Tied \\
\textsc{bbodyrad} & $T_{\text{bb}}$ (keV) & $0.49279\,\dagger$ & Untied \\
\textsc{bbodyrad} & \Nbb & 8857.68 & Frozen \\
\textsc{gauss} & \Eline (keV) & $6.40\pm0.05$ & Tied \\
\textsc{gauss} & $\sigma_\text{line}$ (keV) & 0.82 & Frozen \\
\textsc{gauss} & \Nline & $1.42\pm0.08\times10^{-2}$ & Tied \\
\bottomrule
\end{tabular}
\caption{Values of the best-fitting parameterisation for model 2, with symbols and errors the same as in Table \ref{tab:1BB-bestvals}. \textsc{xspec} returned a fit statistic of $\chis=128.4$ for 500 d.o.f.
\label{tab:2BB-bestvals}}
\end{table}

\subsubsection{Spectral model 3} \label{sec:spec-model-3}

We also considered a different radial temperature dependence or temperature profile of the disc. 
Typically for a standard disc, and assumed by \textsc{diskbb}, the radial temperature dependence of the observed multicolour disc blackbody is $T(r) \propto r^{-p}$ where $p=0.75$. In the \textsc{diskbb} model component, this value of $p$ is fixed. 
However, as discussed in \citet{KubotaMakishima04} and \citet{Kubotaetal05}, general relativistic effects, electron scattering, and/or advection can give a different value of $p$.  Heating by the power-law component may also cause the disc blackbody to deviate from the standard temperature profile.
So, for a varying temperature dependence caused by, e.g. a varying illumination pattern, we tested the model \textsc{phabs}\,$\times$\,\textsc{(simpler\,*\,diskpbb\,+\,gauss)}. In the component \textsc{diskpbb}, the exponent of the radial dependence of the disc temperature $p$ is a free parameter. 
Since \textsc{diskbb} is a multicolour blackbody, and $p$ changes how strongly the spectrum weights the component blackbodies, a smaller $p$ increases the effect of the inner disc radii blackbodies on the total disc blackbody spectrum.
Using \textsc{diskpbb} also affects values of \Fscatt: since the spectrum of \textsc{diskpbb} is flatter than \textsc{diskbb} for $p<0.75$, i.e. there are more lower energy photons, the Comptonising component modelled by \textsc{simpler} must scatter a smaller fraction of the seed spectrum 
to produce the same shape as in models 1 and 2.
For all fits, parameters were given the same value ranges as in model 1, and $p$ was allowed the default range. 

The starting $\chis/$\,d.o.f. was $7113.9 / 568$ with all parameters tied. As with the first two models, untying \Fscatt\ and then $\Gamma$ gave the most significant improvements to the fit ($\chis = 387.3$ for 545 d.o.f., and $\chis=167.0$ for 522 d.o.f., respectively). 
The next parameter that, when untied, gave the most significant improvement to the fit is $p$, returning $\chis=127.9$ for 499 d.o.f. 
Supervising the fitting, it became clear that reducing $p$ effectively lowered \TdiskP\ and \Fscatt.
We note that untying \Nline\ instead of $p$ did not reduce the \chis\, contrary to what was seen for models 1 and 2. Untying additional parameters did not significantly improve the fit. For the final parameterisation, $\sigma_\text{line}$ was frozen at 0.76 and \NdiskP\ was frozen at 560.907 (the best-fitting values) following the same motivation as with models 1 and 2.
A summary of the best parameterisation for model 3, with the value of each parameter, is shown in Table \ref{tab:pBB-bestvals}. 

\begin{table}  
\centering
\begin{tabular}{l l l l}
\toprule
Component & Parameter & Value & Notes \\
\midrule
\textsc{phabs} & $N_\text{H}$ ($\times 10^{21}$~cm$^{-2}$) & 6.0 & Frozen \\
\textsc{simpler} & $\Gamma$ & $2.44\,\dagger$ & Untied \\
\textsc{simpler} & \Fscatt & $0.130\,\dagger$ & Untied \\
\textsc{simpler} & $U$ & 1 & Frozen \\
\textsc{diskpbb} & \TdiskP (keV) & $0.947\pm0.003$ & Tied \\
\textsc{diskpbb} & $p$ & $0.507\,\dagger$ & Untied \\
\textsc{diskpbb} & \NdiskP& $560.907$ & Frozen \\
\textsc{gauss} & \Eline (keV) & $6.49\pm0.05$ & Tied \\
\textsc{gauss} & $\sigma_\text{line}$ (keV) & 0.76 & Frozen \\
\textsc{gauss} & \Nline & $1.23^{+0.07}_{-0.06}\times10^{-2}$ & Tied \\
\bottomrule
\end{tabular}
\caption{Values of the best-fitting parameterisation for model 3, with symbols and errors the same as in Table \ref{tab:1BB-bestvals}. \textsc{xspec} returned a fit statistic of $\chis=126.8$ for 501 d.o.f.
\label{tab:pBB-bestvals}
}
\end{table}

\begin{figure}
\centering
\includegraphics[width=0.99\columnwidth]{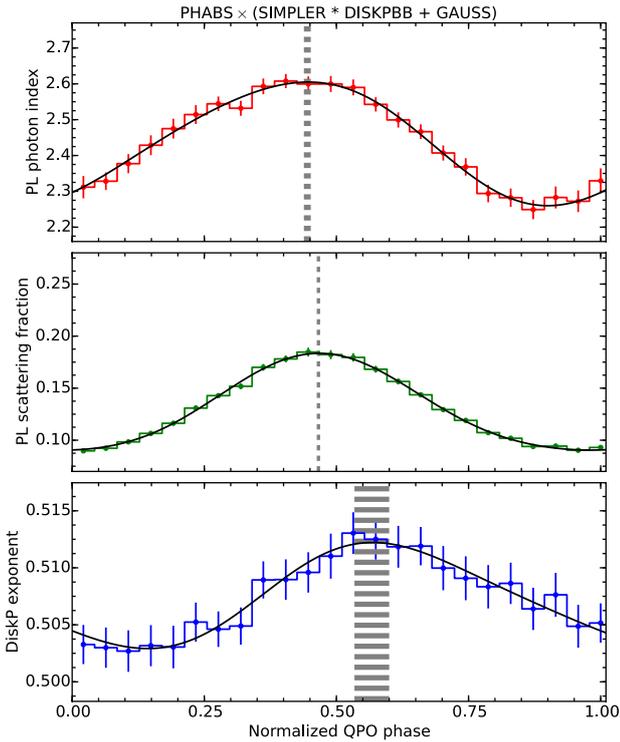}
\caption{
Same meaning as Figure \ref{fig:parvar-1BB}, for model 3.
The top panel shows $\Gamma$ in red, the middle panel shows \Fscatt\ in green, and the bottom panel shows $p$ in blue. 
The fractional rms amplitude variation of $\Gamma$ is 5.0 per cent and \Fscatt\ is 26 per cent.
The normalised phase difference between \Fscatt\ and $\Gamma$ is $0.02 \pm 0.01$, and between \Fscatt\ and $p$ is $-0.10 \pm 0.04$. 
}
\label{fig:parvar-pBB}
\end{figure}

For model 3, the untied parameters are $\Gamma$ (the power-law photon index), \Fscatt\ (the power-law scattering fraction), and $p$ (the exponent of the disc's radial temperature dependence). 
The value of the parameters in each phase bin is shown in Figure \ref{fig:parvar-pBB}. 
Comparing the values of the untied parameters in this model with the previous two models, $\Gamma$ has the same value as in model 2 and is lower than in model 1, \Fscatt\ is lower here than in models 1 and 2, \TdiskP\ is the same as \Tdisk\ in model 2 and is higher than the mean \Tdisk\ in model 1, and \NdiskP\ is much lower than \Ndisk\ in models 1 and 2. 
These differences in the values are consistent with what we expect from the model component \textsc{diskpbb} with $p<0.75$.
The iron line equivalent width varied from $0.21 - 0.29$\,keV.

\subsection{Parameter variation fitting} \label{sec:bootstrapping}

In order to study how the spectral parameters inferred from the CCF vary and to what significance, we must account for any error that is introduced due to the spectral data not being independent in time. Since the spectral fitting software assumes that the simultaneously fit data and errors on the data are fully independent, we cannot trust the errors from our spectral analysis. Therefore, to estimate the accuracy of the untied parameter variations, we implemented bootstrapping from our data set.
Bootstrapping requires selecting (with replacement) $M$ times from $M$ segments of data.
For our data set, 198 segments requires 5537 bootstrap iterations to converge on the true error distribution.\footnote{Using $M\cdot (\ln(M))^2$ \citep{FeigelsonBabu}. } 
Each bootstrap iteration consisted of selecting CCF segments to make a new average energy-dependent CCF, through to the phase-resolved spectroscopy.  Each bootstrapped average energy-dependent CCF was used to determine a set of phase-binned spectral parameters for each spectral variability model, in the same way as for the data.  The standard deviations for each parameter were determined from the distributions obtained from all bootstrap iterations and are used to determine the errors plotted in Figures~\ref{fig:parvar-1BB}, \ref{fig:parvar-2BB} and \ref{fig:parvar-pBB}.

Even though the untied spectral parameters are free to have any value within the physically allowed ranges given previously, and despite the apparently complex form of spectral variability seen in Figure~\ref{fig:ccf-spec}, the parameters appear to vary in a roughly sinusoidal fashion. 
The function fitted to the variations is
\begin{equation} \label{eqn:fitfunc}
y(t) = A_1 \sin\left(2\pi t + \phi_1\right) + A_2 \sin\left(4\pi t + \phi_2\right) + D\,,
\end{equation}
where $A_1$ and $A_2$ are amplitudes, $\phi_1$ and $\phi_2$ are phase offsets, and $D$ is a $y$-axis offset. The function is fixed so that the second part is a harmonic, with twice the frequency of the first part. We refer to this function as a whole as the ``fit function''. The fit parameters $A_1$, $A_2$, $\phi_1$, $\phi_2$, and $D$ were found for each bootstrap iteration by a least-squares fit (using scipy.optimize.leastsq) to the untied parameter value variations.

If there is a strong harmonic present in the QPO data, the CCF erases any phase offset between the fundamental and the harmonic. This is because the CCF picks out only the \textit{relative} offset of the channel-of-interest fundamental and harmonic with the reference band fundamental and harmonic respectively. So, any harmonic content is added into the CCF such that its phase is relative to the reference band phase of the harmonic and thus the phase dependence of spectral parameters need not be aligned with those of the fundamental. The GX 339--4 data used in this analysis has a weak harmonic in the 3--5 keV power spectrum, but more harmonic content at higher energies (Figure \ref{fig:psd}). We do not exclude the possibility that the stronger harmonic at higher energies could give a harmonic in the untied parameter values, and so we included a harmonic with all parameters free in the fit function. Overall we found that $A_2$ is at least a factor 4 times smaller than $A_1$ for all variable parameters and is only formally significant for \Fscatt.

\subsubsection{Phase relationships of the spectral parameters}\label{sec:dphase}

\begin{table*}  
\centering
\begin{tabular}{l l c c c c c c}
\toprule
Model & Parameter & $A_1$ & $\phi_1$ & $A_2$ & $\phi_2$ & $D$ & $\phi_\text{max}$  \\
\midrule
 & $\Gamma$ & $0.15\pm0.01$ & $0.834\pm0.009$ & $0.02\pm0.02$ & $0.1\pm0.3$ & $2.518\pm0.008$ & $0.449\pm0.010$ \\
1 & $\Fscatt$ & $0.053\pm0.003$ & $0.792\pm0.004$ & $0.007\pm0.001$ & $0.48\pm0.02$ & $0.157\pm0.001$ & $0.454\pm0.005$ \\
 & $\Tdisk$ & $0.0041\pm0.0004$ & $0.16\pm0.02$ & $0.001\pm0.001$ & $0.8\pm0.7$ & $0.831\pm0.002$ & $0.13\pm0.02$ \\
\midrule
 & $\Gamma$ & $0.17\pm0.02$ & $0.859\pm0.005$ & $0.01\pm0.01$ & $0.1\pm0.2$ & $2.44\pm0.01$ & $0.417\pm0.007$ \\
2 & $\Fscatt$ & $0.059\pm0.003$ & $0.811\pm0.002$ & $0.007\pm0.001$ & $0.892\pm0.008$ & $0.180\pm0.003$ & $0.436\pm0.002$ \\
 & $\Tbb$ & $0.0060\pm0.0006$ & $0.12\pm0.01$ & $0.001\pm0.001$ & $0.2\pm0.2$ & $0.493\pm0.002$ & $0.16\pm0.02$ \\
\midrule
 & $\Gamma$ & $0.17\pm0.01$ & $0.851\pm0.006$ & $0.02\pm0.01$ & $0.1\pm0.2$ & $2.44\pm0.01$ & $0.425\pm0.008$ \\
3 & $\Fscatt$ & $0.047\pm0.002$ & $0.800\pm0.003$ & $0.006\pm0.001$ & $0.37\pm0.01$ & $0.131\pm0.001$ & $0.446\pm0.004$ \\
 & $p$ & $0.0045\pm0.0005$ & $0.67\pm0.02$ & $0.001\pm0.001$ & $0.3\pm0.5$ & $0.507\pm0.002$ & $0.55\pm0.04$ \\
 \bottomrule
\end{tabular}
\caption{ Values of each variable of the fit function (Equation \ref{eqn:fitfunc}) for the untied parameters in each spectral model. $\phi_\text{max}$ is the phase of the maximum value of the fit function (indicated with the dashed gray lines in Figures \ref{fig:parvar-1BB}-\ref{fig:parvar-pBB}). 
The errors are computed as the standard error from bootstrapping. The values of $\phi$ have been normalised to the range 0 to 1.
\label{tab:fitfuncval}
}
\end{table*}

The parameters of the fit function for each untied parameter per model are shown in Table \ref{tab:fitfuncval}.
The phase differences between the untied parameters were measured from the maximum of the fit function $\phi_\text{max}$ for each parameter, with the errors corresponding to the standard deviation obtained from bootstrapping. 
Since \Fscatt\ was the best-constrained untied parameter, we computed the phase difference of the two other untied parameters with respect to it.  
The difference in phase between the variations in the untied parameter values are also quoted in Table \ref{tab:specmodels}. 
The phases of \Fscatt\ and $\Gamma$ are very close: $0.01 \pm 0.02$, $0.019 \pm 0.009$, and $0.02 \pm 0.01$ for models 1, 2, and 3, respectively (normalised to the range 0 to 1; 1-$\sigma$ errors). 
The phase difference for model 1 between \Fscatt\ and \Tdisk\ is $\Delta\phi = 0.32 \pm 0.02$.
For model 2, \Fscatt\ and \Tbb\ and  have $\Delta\phi = 0.28 \pm 0.02$. 
For model 3, \Fscatt\ and $p$ are out of phase by $\Delta\phi = -0.10 \pm 0.04$. As previously noted, lowering $p$ gives the same effect as raising \TdiskP\ and \Fscatt, so these results are consistent with the previous two models finding that the blackbody increases before the power-law increases.  
If we instead compare the phase of the maximum of \Fscatt\ with the phase of the \textit{minimum} of $p$, we see a phase difference of $0.33 \pm 0.02$, consistent with the phase lead seen for the blackbody temperature variations in models 1 and 2.
In all three models, the blackbody parameter values have a small variation, but as evidenced by the systematic spectral analysis, they are required to vary. 
Before considering the physical interpretation of the results in Section \ref{sec:discussion}, we first checked that the QPO spectral parameter variations inferred from our CCF method can also reproduce the shape of the lag-energy spectrum.

\subsection{Comparison with the lag-energy spectrum} \label{sec:lags}
The last step of the analysis was to use the best-fitting energy spectral models to simulate lag-energy spectra and compare them to the data. 
This was done as a secondary check on the models in the Fourier domain, where the QPO can be selected directly in frequency and the errors are better-behaved.

We selected untied parameter values from the smooth fit function at each of the 24 QPO phases, and simulated a PCU2 event-mode spectrum for each phase bin using the \textsc{xspec} command \texttt{fakeit}, with the same instrumental response matrix as the data. 
The spectra are simulated without Poisson errors, to provide an idealised model. 
We then tiled the spectra to make a light curve with the same length as a segment used to calculate the CCF, and turned the spectra into photon count-rate light curves in each energy channel. 
From here, the simulated models were treated in the same way as the data for lag-energy spectral analysis.

\begin{figure}
\centering
\includegraphics[height=0.99\columnwidth,angle=90]{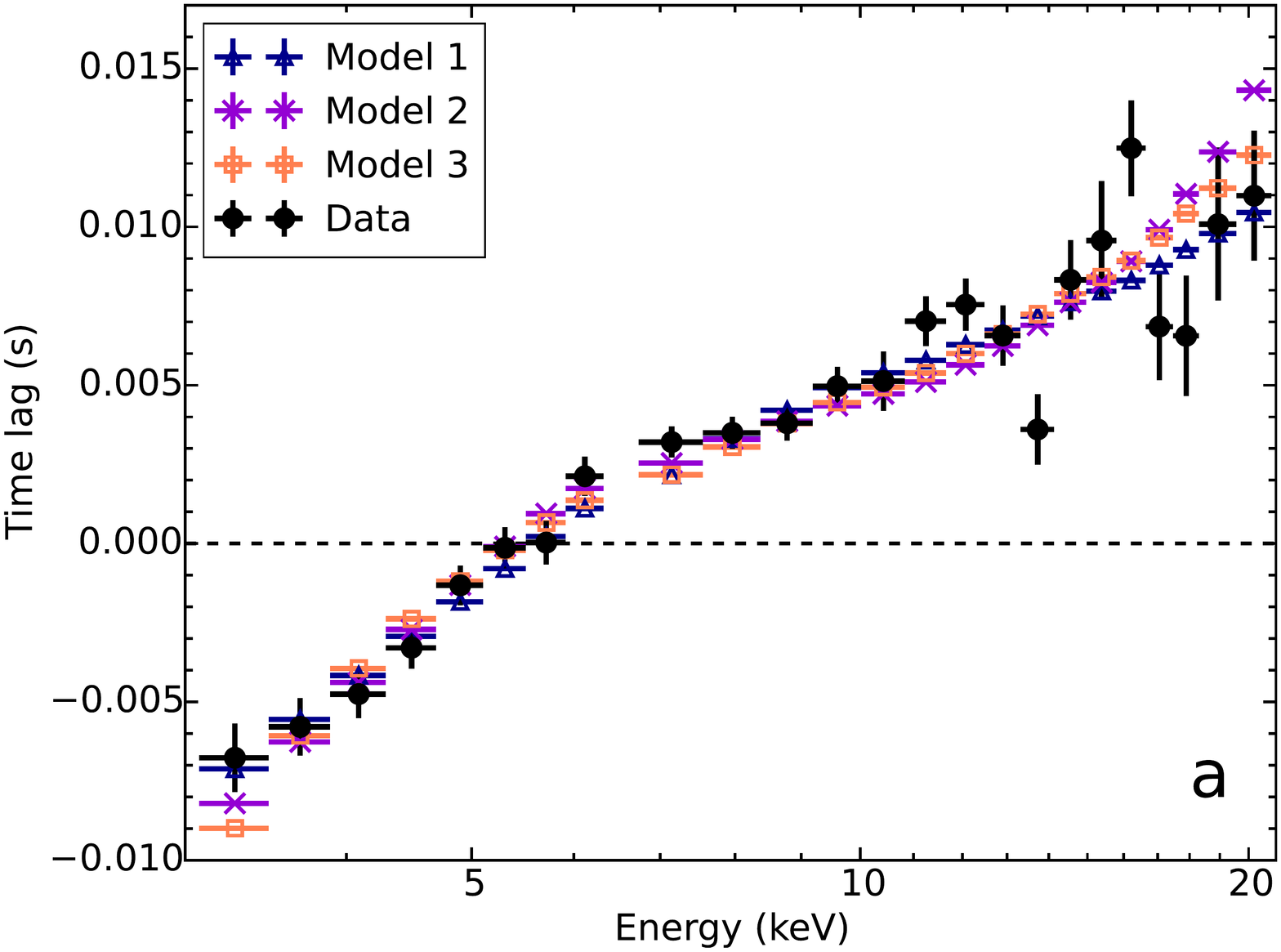}
\includegraphics[height=0.99\columnwidth,angle=90]{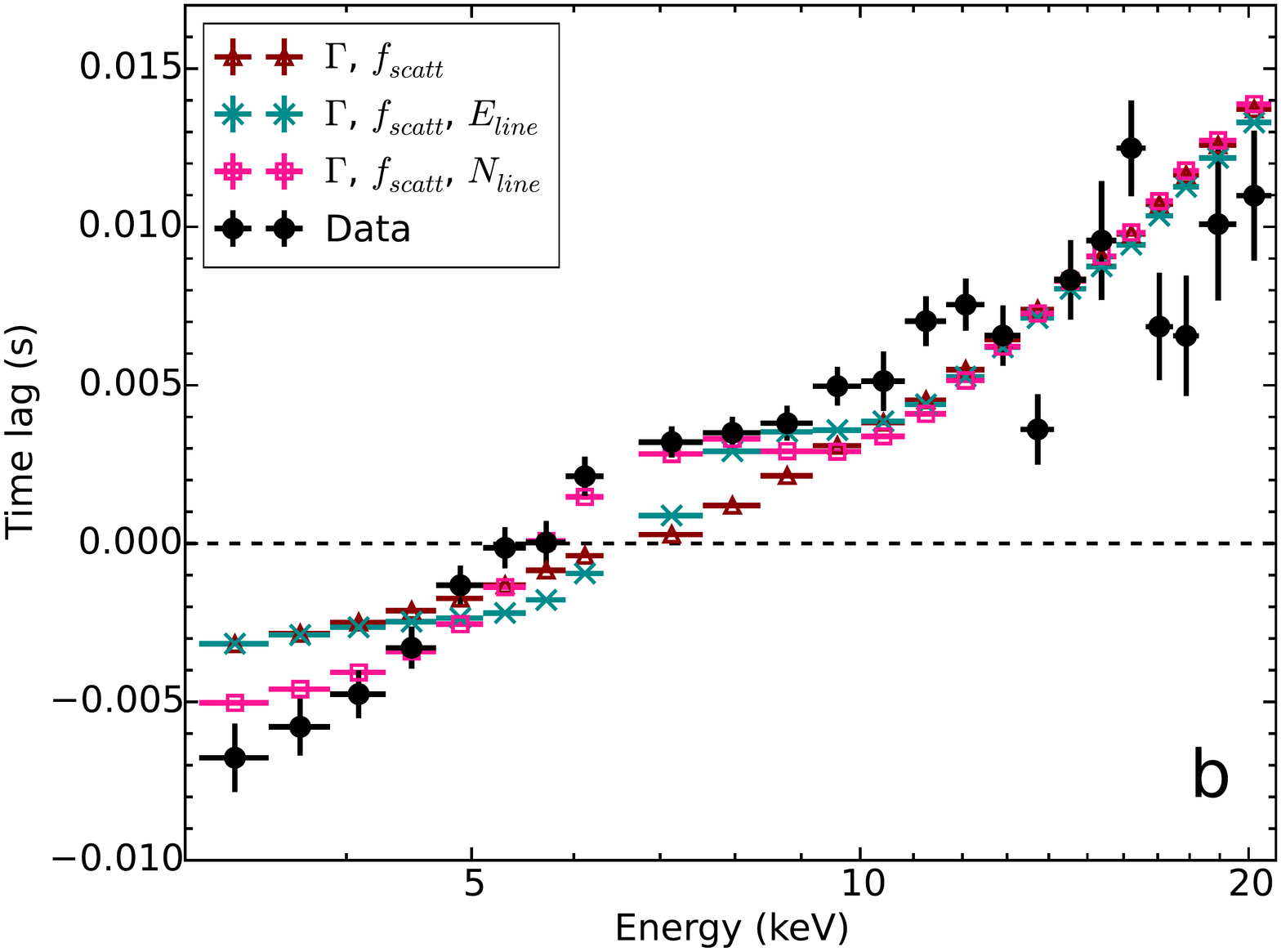}
\caption{ 
Lag-energy spectra obtained from the cross-spectrum averaged over 4--7 Hz.
The black dots are the data (same as in Figure \ref{fig:lag-energy}).
The black dashed line indicates no time lag with respect to the variability in the reference band.
\textbf{a}:
The dark blue triangles are simulated using model 1 ($\chis = 37.9$). 
The purple X's are simulated using model 2 ($\chis= 45.5$).
The orange squares are simulated using model 3 ($\chis = 47.2$).
\textbf{b}: These simulations are alternative parameterisations of model 1 that fit the data poorly.
The dark red triangles are simulated with $\Gamma$ and \Fscatt\ untied ($\chis= 175.6$).
The dark cyan X's are simulated with $\Gamma$, \Fscatt, and \Eline\ untied and $\sigma_\text{line}=0.97$ ($\chis = 149.1$).
The pink squares are simulated with $\Gamma$, \Fscatt, and \Nline\ untied and $\sigma_\text{line}=0.97$ ($\chis = 82.1$).
All of the model-comparisons in panels \textbf{a} and \textbf{b} have 24 degrees of freedom.
\label{fig:lag-energy-comparison} }
\end{figure}

In Figure \ref{fig:lag-energy-comparison}a we show simulated lag models plotted with the data from Figure \ref{fig:lag-energy} (model 1, in dark blue, gives $\chis = 37.9$; model 2, in purple, gives $\chis = 45.5$; model 3, in orange, gives $\chis =47.2$; all have 24 d.o.f.).  
While all three of the QPO spectral models can match the overall `broken' shape of the lag-energy spectrum quite well, models 2 and 3 slightly deviate from the slope of the observed lag-energy spectrum at low ($<$\,4\,keV) and high ($>$\,16\,keV) energies.

For comparison, in Figure \ref{fig:lag-energy-comparison}b, we also plot the simulated lag-energy spectra from some alternative parameterisations of model 1. 
The first, in dark red, has only \Fscatt\ and $\Gamma$ untied. 
It is clear from the plot that this model cannot re-create the slope below $5\,$keV or the break in the lag at $\sim$6.5\,keV. 
Likewise, the second and third parameterisations vary \Eline\ and \Nline, respectively, in addition to \Fscatt\ and $\Gamma$ and with the iron line width frozen at $\sigma_\text{line} = 0.97$. 
Neither can reproduce the slope below $5\,$keV, and the second cannot make a bump with the right shape at the right energy. 
All three of these lag models have the same high-energy slope as model 2 from panel a. These poor matches are reflected in the fits to the data: $\chis = 175.6$, $\chis = 149.1$, and $\chis=82.1$, respectively, each for 24 degrees of freedom.

The bootstrapping technique accounts for statistical error, but it is possible that some systematic error is unaccounted for. However, since simulations from the bootstrapping-verified parameter variations can reproduce the lag-energy spectrum of the data well, our phase-resolved spectroscopy technique does not seem to introduce a systematic bias to the spectral-timing properties of the data.

The lag-energy spectra simulated from the QPO phase-resolved spectral models further support a spectral model for QPO variability consisting of a varying blackbody preceding (by relative phase $\sim$\,0.3) a varying power-law, however the lags show a better fit with model 1 (as opposed to the phase-resolved spectral fits, which preferred models 2 and 3). 
As demonstrated here, it is very important to account for the Fourier-domain cross spectra as well as the time-domain energy spectra when fitting a model to the data, as both spectral and timing information are pertinent. 
Furthermore, our results show that it is possible to reproduce the lag-energy spectrum of a Type B QPO by sinusoidally varying spectral parameters with different relative phases.

\section{Physical interpretation} \label{sec:discussion}
We fitted three spectral models to phase-resolved spectra of the Type B QPO from the 2010 outburst of GX 339--4 which we determined using our CCF method. 
All of our models consisted of a blackbody-like disc component and a power-law component produced by Compton up-scattering of the disc blackbody photons, but parameterised the blackbody contribution in different ways: either variations of an entire multiple-temperature disc blackbody, variations of a single-temperature blackbody on top of a constant disc blackbody, or variations in the radial temperature profile of a multi-temperature disc blackbody. 
All three models showed that the blackbody must vary as well as the power-law on the QPO timescale: the varying blackbody temperature leads the varying power-law scattering fraction \Fscatt\ by a relative phase of $\sim 0.3$, and the varying power-law photon index $\Gamma$ is very close in phase to the variations in \Fscatt\ (with a model-dependent relative phase lead of up to $\sim 0.02$).  
These results are checked with and supported by bootstrapping analysis and simulated lag-energy spectral models compared with the data. 

In terms of variability amplitudes, the power-law component dominates the QPO variability, showing fractional rms variations in scattering fraction of $\sim25$~per~cent, when calculated directly from the sinusoidal fit function applied to each model.  
In comparison, the blackbody temperature in model~1 shows a fractional rms of only 0.35~per~cent, corresponding to a flux variation of only 1.4~per~cent.  
The total blackbody flux variation is similar for the other models (e.g. after including dilution of the larger single blackbody rms variability by the constant disc blackbody contribution in model 2).
These results are consistent with the findings of \citet{Gaoetal14} that the power-law dominates the Type B QPO emission in GX 339--4 (and, more generally, is consistent with the findings of \citealt{RemillardMcClintock06,SobolewskaZycki06} and \citealt{Axelssonetal14} for LFQPOs).  
However, despite the dominant role of the power-law emission in the observed QPO variability, the large phase-lag of the power-law relative to the blackbody has a significant effect on the lag-energy spectrum, causing the distinctive `break' feature, as shown in Figure~\ref{fig:lag-energy-comparison}.

Although our spectral models allowed for different types of blackbody variability, all point to a very similar phenomenological picture of the QPO spectral variability. 
The parameter variations show large changes in the power-law index and normalisation (measured by $\Gamma$ and \Fscatt) and small changes in the blackbody emission. 
For model 1, the \Tdisk\ variations represent the entire disc blackbody modestly varying in effective temperature, suggesting that it is not the hottest region of the disc which varies.  
This picture is supported when we keep \Tdisk\ constant with phase in model 2 and allow a separate single blackbody component to vary, resulting in modest variations of the single blackbody $T_\text{bb}$ but with a {\it lower} temperature than the inner disc temperature. 
Allowing the disc radial temperature profile to vary in model 3 also supports this view, with the variations in profile mimicking the variations in the disc spectrum down to lower energies that are seen in the other two models.

We now consider the broad physical implications of our results.  
The simplest explanation for linked blackbody and power-law variability is that the blackbody drives the power-law variation by varying the seed-photon flux which illuminates the Compton scattering region.  
This interpretation suffers from the problem that the observed phase lag corresponds to a light-travel time of $\sim$\,60\,ms or a distance of $\sim$\,1.8$\times10^4$\,km: $\sim$\,1700\,\rg for a $7\Msun$ black hole.  
This distance is much larger than the disc inner radius inferred from the disc blackbody normalisation of $\sim$100--260~km\footnote{Here we assume a disc spectral hardening factor $f\sim1.7$ (see \citealt{ShimuraTakahara95}), inclination 40~degrees \citep{MunozDariasetal13} and calculate values for the conservative distance range of 6--15\,kpc \citep{Hynesetal04}.}, so any model seeking to explain the lag in terms of light-travel times must assume an unfeasibly large height of the Compton-scattering region above the disc.  
Furthermore, if the observed photon index fractional rms variation of $\sim$\,5~per~cent is produced by changes in the seed to heating luminosity ratio in the Compton scattering region, the required variations in seed luminosity would need to be much larger than the observed disc variability (e.g. as much as $\sim$\,30~per~cent, applying the equation from \citealt{Beloborodov99}).

Another possibility is that quasi-periodic variations in accretion rate are produced in the accretion disc and propagate inwards to produce a lagging power-law component from an inner corona, as seems to be the case for the hard state broadband noise variability in GX~339-4 and other black hole systems \citep{Uttleyetal11}.  
This interpretation also seems unlikely, since for the Type~B QPO the disc blackbody variations are much smaller than the power-law variations (unlike the situation for the broadband noise).  
It might be possible that the accretion rate variation originates only in the innermost radii of the disc and so modulates only a small fraction of the blackbody emission, but then the lag would likely be washed out by variable heating effects of the strong power-law variations (see below).

An alternative possibility is that the fluctuations in blackbody emission are driven by X-ray heating from intrinsically varying power-law emission.  
According to model~1, the average total power-law flux is between 40 and 60~per~cent of the total disc blackbody flux (assuming a power-law high-energy cut-off $>20$~keV).  
If the power-law emitting region is isotropically emitting and the disc albedo is 0.3 (e.g. \citealt{GierlinskiDonePage08}), then given the observed 25~per~cent power-law rms variation, the corresponding disc rms variability should be (0.07--0.1)$\times f_{\rm cov}$, where $f_{\rm cov}$ is the fraction of the sky covered disc as seen from the power-law.  The observed disc variability could then easily be produced by X-ray heating even for modest covering fraction of $\sim 0.2$.  
However, we then would expect the blackbody variations to be in phase with those of the power-law, or at least lag by a (small) light-travel time, which is not the case.  
Therefore we are left with the simplest explanation for our results, which is that the variation in power-law emission is not intrinsic but {\it geometric}, in which case the observer and the disc need not see the same power-law emission, and large phase lags can be produced by the varying system geometry.  
This interpretation is consistent with the idea that the inner power-law emitting region is precessing, perhaps via the Lense-Thirring mechanism (e.g. \citealt{StellaVietri99}). 
We will therefore consider this scenario in more detail. 
However, we do not rule out the possibility that other models not considered here can explain the observed energy spectral changes, and we stress again that the phase-resolved spectroscopy technique presented in this paper is applicable regardless of the physical interpretation of the results.

First, we should consider that as seen from the disc, a precessing power-law emitting region should (quasi-)periodically illuminate different azimuthal regions of the disc, producing a local enhancement of the blackbody emission due to X-ray heating.  
Even if the power-law emission is intrinsically constant, an observer will see periodic blackbody variations as a result, due to the Doppler boosting and deboosting of the region of enhanced blackbody emission as the power-law emitting region illuminates the approaching and receding sides of the disc as seen by the observer.

Type C QPOs, which occur on timescales very similar to Type Bs but are observed across a wider range of outburst states, might arise from Lense-Thirring precession of a hot inner flow inside the inner edge of the accretion disc \citep{FragileAnninos05, IngramDoneFragile09, Axelssonetal14, IngramvdKlis15, Fragileetal16}. 
A hot inner flow would have a small scale height, and the precession would effect more significant changes in illumination and heating of the inner edge of the accretion disc than in the cooler parts of the disc. 
Since the inner edge of the accretion disc shows the strongest Doppler boosting, the varying illumination pattern should show a large blackbody modulation.  
For example, at a disc radius of 10~\rg and assuming a disc inclination of 40~degrees, the maximum Doppler factor for gas on the approaching side of the disc is 1.13, so that allowing for boosting and deboosting, the peak to trough apparent temperature shift should be 27~per~cent compared with 2.4~per~cent observed for model 2 (which shows the most extreme temperature change).  
Even a large emitting radius in the disc of 100~\rg will produce a peak-to-trough shift of 10~per~cent.  
However, this variation assumes that only a small range of azimuths on the disc is illuminated, to maximise the observed temperature shift.

The problem can therefore be solved by assuming that the illuminated region of the disc is large and only undergoes small shifts due to precession, such that the observed Doppler boost is an average over many azimuths and changes only by a small amount in response to the changing illumination pattern.  
Such an effect might be obtained if the precessing power-law emitting region which produces the Type B QPO has a relatively large scale height compared to the inner disc radius (up to tens of \rg), i.e. it is more {\it jet-like} than disc-like (as for the Type C QPOs).  
This picture is consistent with the findings of \citet{Mottaetal15}, that the Type B QPO integrated rms is larger for more face-on binary systems and the Type C QPO integrated rms is larger for more edge-on systems, implying that the emission geometries are different. 

\begin{figure}
\centering
\includegraphics[width=0.85\columnwidth]{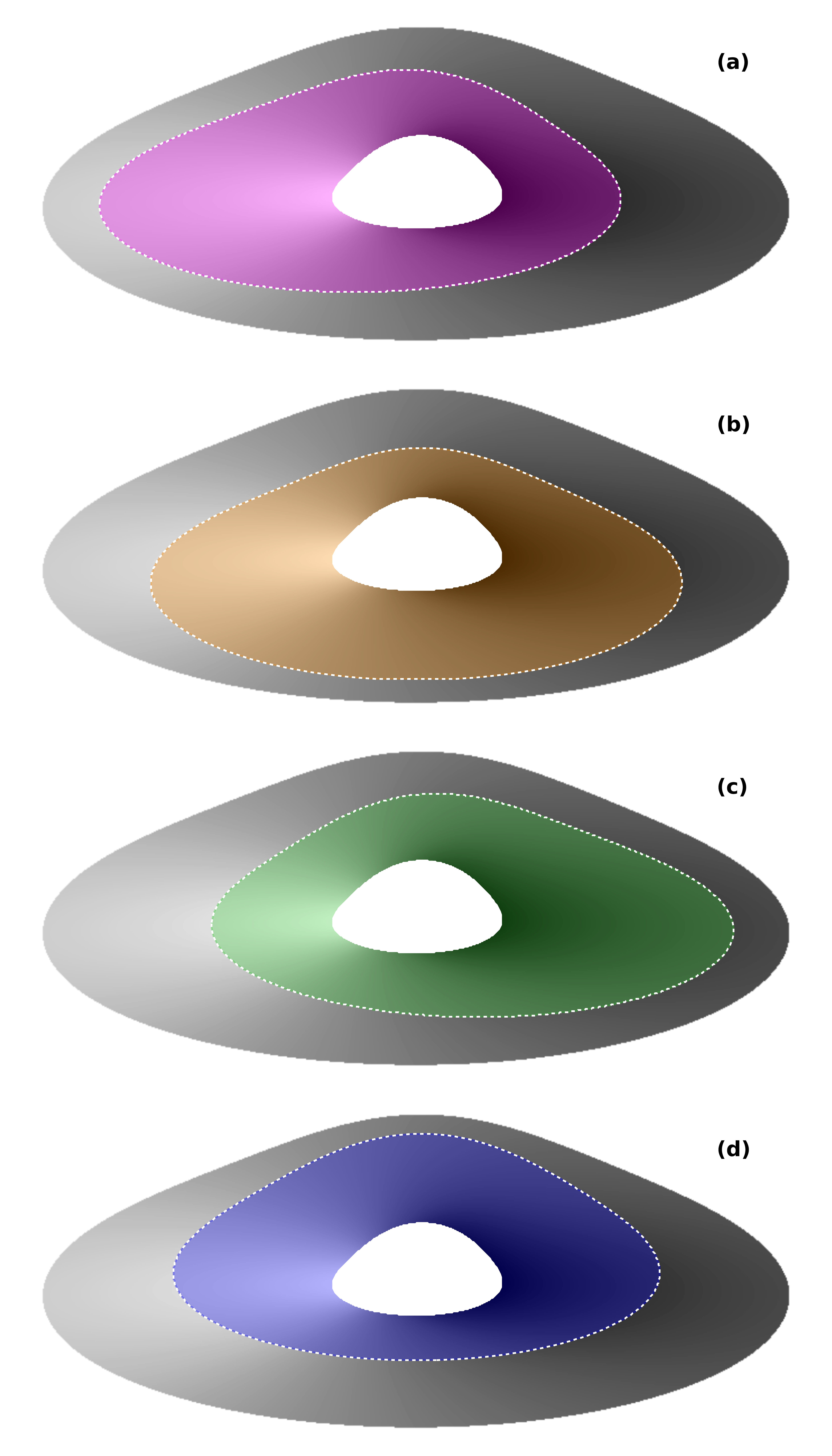}
\caption{A simplified depiction of the illumination footprint on the inner portion of the accretion disc by a large-scale-height precessing Comptonising region as it sweeps around. 
The grayscale colour-mapping of the disc shows the intensity of the emission; light shading means higher observed intensity (which varies as a function of Doppler blueshift and intrinsic emission). 
The accretion flow is orbiting in a counter-clockwise direction. 
The illuminated region of the accretion disc is shown in magenta, orange, green, and blue for phases $a$, $b$, $c$ and $d$ respectively in a QPO cycle (which we assume are each separated by 0.25 in relative phase).}
\label{fig:geometry}
\end{figure}

Detailed modelling of the effects of a variable illumination pattern on the disc blackbody emission is beyond the scope of this work, but we illustrate our interpretation by showing in Figure~\ref{fig:geometry} the illumination footprint on the disc at four key phases (separated by 0.25 in relative phase).  
At phase $a$, the apparent blackbody temperature is at a maximum when the power-law is preferentially illuminating the Doppler-boosted approaching side of the accretion disc, while at phase $c$ the opposite, deboosted side of the disc is preferentially illuminated, leading to a minimum in apparent temperature.  
Linking the power-law emission to the illumination pattern is less-clear since it depends on the orientation-dependence of the power-law emission.  
Here we assume that the illumination pattern precesses in the same direction as the orbital motion, so that for a jet-like power-law emitting region, we might expect the jet to point more towards the observer at phase $b$ and more away from the observer at phase $d$.  
Comparing with our observed pattern of disc and power-law parameter variations (e.g. in Figure~\ref{fig:parvar-1BB} for model 1), the peak in power-law emission and photon index would coincide roughly with the time when the jet-like emitting region points towards us at phase $b$, suggesting that Doppler boosting of the emitting region may play a role in the observed power-law flux variability, which might be expected if the emitting region is jet-like in geometry.  

Note that in the above scenario we do not necessarily expect the peak in blackbody emission and power-law emission to be separated by exactly 0.25 in relative phase, which is close to but significantly different from the phase difference we observe.  
The difference will also be influenced by the shape of the illuminating region on the disc and hence the relative weighting of emission from different parts of the disc, and we would also anticipate that light-bending effects should play a role.  
Such effects can be predicted in future using ray-tracing simulations.  
Our toy model also does not simply explain the tight relation (and very small phase offset) between the scattering fraction (or effectively, power-law normalisation) and photon index, although this result suggests that the spectral shape of the power-law emission is also orientation-dependent.  
Also we do not observe the variations in the iron line energy (or flux) on the QPO timescale that are predicted by \citealt{FragileMillerVandernoot05} and \citealt{IngramDone12} for LFQPOs produced by precessing hot flows in X-ray binaries. 
Since the disc blackbody varies by only a very small fraction, if this effect were due to variable illumination combined with Doppler boosting, we would expect a similarly small fractional amplitude of variation in the iron line emission. 
Such a variation would not be detectable given the lack of sensitivity to a very small iron line flux variation and the coarse energy resolution at the expected iron line energy ($5.5-7.0$\,keV).

An obvious interpretation of our toy model geometry is that the jet-like power-law emitting region is actually the base of the larger-scale radio-emitting jets seen in the hard and intermediate states of black hole X-ray binary systems.  
A jet-base origin for the X-ray power-law emission in X-ray binaries was first proposed by \citet{MarkoffNowakWilms05}.  
A number of works support the possibility of a connection between Type B QPOs and jets.  
Type B QPOs are only seen at the crossover between the soft and hard spectral states, in the soft intermediate state \citep{Belloni10, Heiletal15a}.   
During this transition in previous outbursts, there are observations of radio flaring and evidence of the radio-emitting jet switching off \citep{FenderBelloniGallo04, FenderHomanBelloni09}.  
Moreover, \citet{Kalamkaretal16} have reported the discovery of an infra-red LFQPO in GX~339-4, coincident with an X-ray type C QPO at the first harmonic of the IR QPO.  
Their results suggest that low-frequency quasi-periodic phenomena such as precession could extend to or affect the IR--producing part of the jet, which would make it plausible that precession also affects the X-ray emitting jet base in the case of the Type B QPO.

\section{Conclusions}
We have demonstrated a new spectral-timing technique for phase-resolved spectroscopy using the cross-correlation function, to enable deeper understanding of the QPO mechanism in black holes and neutron stars. 
The technique allows us to quantitatively probe how the spectrum of a source changes as a function of QPO phase.  
For the Type B QPO in GX 339--4's 2010 outburst observed with RXTE, the spectral shape changes on the QPO timescale. 
Specifically, we find that the power-law normalisation has a $\sim$\,25~per cent fractional rms amplitude variation, the power-law photon index has $\sim$\,5~per cent rms variation, while the blackbody varies significantly, but with an rms variation in flux of only 1.4~per cent.  
Crucially, the varying blackbody-like component leads the varying power-law component by $\sim$\,0.3 of a QPO cycle.   
This combination of large power-law variability with small blackbody variability, combined with a large phase lag of the power-law relative to the blackbody variation, implies that the QPO flux variations are not primarily intrinsic to the emitting regions but geometric.  
The spectral and flux variability and lag suggest a large-scale-height Comptonising region (such as the base of a jet) which illuminates and heats different but overlapping regions of the inner accretion disc as it precesses during a QPO cycle. 

Regardless of the physical interpretation, our method may be used as a guide to understand quasi-periodic variability in terms of the different spectral components.  
The statistical errors across phase-bins are not statistically independent however, so caution should be applied when interpreting the detailed results, with bootstrapping or other Monte Carlo methods applied to determine uncertainties on spectral parameter variations.  
Furthermore, the spectral component variability inferred from our method should always be compared with statistically robust spectral-timing products, e.g. derived from the cross-spectrum.  
Looking further ahead, more sophisticated fitting techniques (e.g. using the error covariance between phase bins) could be developed to carry out more robust fitting of the phase-resolved spectra.  
Alternatively, direct frequency-domain fitting of the QPO cross-spectrum offers be a simple and robust approach, but may suffer from the problem that variable spectral modelling in the frequency domain is not intuitive, compared to time-domain fitting which may be more easily related to conventional spectral-fitting of time-resolved spectra.

Our results highlight the power of the combination of high count-rates and improved spectral and timing resolution in understanding the innermost regions of accreting compact objects.  
Unfortunately, telemetry constraints limited {\it RXTE} PCA observations of brighter sources than GX~339--4 to data modes with even coarser energy binning than the event mode data we use here.  
However, the recently launched {\it ASTROSAT} mission \citep{ASTROSAT} suffers no such constraints on data from its Large-Area X-ray Proportional Counter (LAXPC) experiment, which offers full event telemetry for much brighter sources (and with larger effective area than the PCA).  
Similarly, we anticipate important breakthroughs from applying these techniques to data from the forthcoming {\it Neutron Star Interior Composition Explorer} ({\it NICER}, \citealt{NICER}), since {\it NICER} will obtain the first data with a soft X-ray response and CCD-level spectral-resolution for Crab-level and higher fluxes, without significant deadtime or instrumental pile-up effects.  
Thus {\it NICER} will allow us to probe the quasi-periodic variability of Fe~K shape due to its superior energy resolution, and distinguish between different models for blackbody variability, thanks to its soft X-ray response.

Since the signal-to-noise of X-ray binary spectral-timing measurements scales linearly with count rate \citep{Uttleyetal14}, future large-area X-ray observatories with dedicated timing capability and large ($>$\,few m$^{2}$) collecting areas at iron line energies, will enable our technique to perform detailed phase-resolved tomography of the Fe K line, even for weak QPO signals.  
In fact, our CCF method was originally conceived to demonstrate tomography of the much weaker HFQPO signatures in data from the proposed {\it Large Observatory for X-ray Timing} ({\it LOFT}, \citealt{Ferocietal12}), which could directly reveal the orbital motion in strong field gravity of emission structures producing the HFQPOs.\footnote{See Figure 2-17 in \url{http://sci.esa.int/loft/53447-loft-yellow-book/\#}} 
Thus the phase-resolved spectroscopy of QPOs offers a powerful probe of the behaviour of matter in the most strongly curved space-times and we encourage further development of spectral-timing techniques to reveal the inner workings of QPOs. 

\section*{Acknowledgements}

A.L.S. acknowledges support from NOVA (Nederlandse Onderzoekschool voor Astronomie).
We thank Adam Ingram, Lucy Heil, Victoria Grinberg, Michiel van der Klis, and the participants of `The X-ray Spectral-Timing Revolution' Lorentz Center workshop (February 2016) for useful discussions that contributed to the development of this paper. We also thank the anonymous referee for their helpful comments.

This research has made use of data and software provided by the High Energy Astrophysics Science Archive Research Center (HEASARC); 
NASA's Astrophysics Data System Bibliographic Services; 
NumPy v1.9.3 and Scipy v0.16.0 \citep{ScipyRef};
Astropy v1.0.4 \citep{AstropyPaper};
Matplotlib v1.4.3 \citep{MatplotlibRef};
iPython v3.2.0 \citep{IPythonRef};
and the AstroBetter wiki.


\bibliographystyle{mnras}
\bibliography{ms_arxiv}  

\appendix
\section{Software}

The software developed for the phase-resolved spectroscopy technique presented in this paper will be made publicly available 6 months after this paper's publication. The repository links are as follows:
\begin{itemize}
\item power spectra: \\ \url{https://github.com/abigailStev/power_spectra}
\item lag spectra: \\ \url{https://github.com/abigailStev/lag_spectra}
\item cross-correlation: \\ \url{https://github.com/abigailStev/cross_correlation}
\item phase-resolved spectroscopy: \\ \url{https://github.com/abigailStev/energy_spectra}
\item \textsc{simpler} \textsc{xspec} model: \\ \url{https://github.com/abigailStev/simpler}
\item simulating spectral-timing data: \\ \url{https://github.com/abigailStev/simulate}
\end{itemize}

\bsp
\label{lastpage}
\end{document}